\documentclass[prl,twocolumn,showpacs,preprintnumbers,amsmath,amssymb]{revtex4}

\usepackage{mathrsfs}
\usepackage{graphicx}
\usepackage{dcolumn}
\usepackage{bm}
\usepackage{comment}

\begin{document}

\title{Unified framework for the entropy production and the stochastic interaction based on information geometry}

\author{Sosuke Ito$^{1,2}$, Masafumi Oizumi$^3$, and Shun-ichi Amari$^4$}
\affiliation{$^1$ Universal Biology Institute, The University of Tokyo, 7-3-1 Hongo, Bunkyo-ku, Tokyo 113-0031, Japan\\$^2$JST, PRESTO, 4-1-8 Honcho, Kawaguchi, Saitama, 332-0012, \\$^3$ The University of Tokyo
3-8-1 Komaba, Meguro-ku, Tokyo 153-8902, Japan, \\$^4$ RIKEN CBS Hirosawa 2-1, Wako-shi, Saitama 351-0198, Japan}
\date{\today}

\begin{abstract}
We show a relationship between the entropy production in stochastic thermodynamics and the stochastic interaction in the information integrated theory. To clarify this relationship, we newly introduce an information geometric interpretation of the entropy production for a total system and the partial entropy productions for subsystems. We show that the violation of the additivity of the entropy productions is related to the stochastic interaction. This framework is a thermodynamic foundation of the integrated information theory. We also show that our information geometric formalism leads to a novel expression of the entropy production related to an optimization problem minimizing the Kullback-Leibler divergence. We analytically illustrate this interpretation by using the spin model.
\end{abstract}

\maketitle
Information geometry~\cite{amari2007methods, amari2016information} is differential geometric theory for elucidating various results in information theory, probability theory and statistics. Applications of information geometry have been found in a variety of fields including machine learning~\cite{amari1992boltzmann}, neuroscience~\cite{amari1995neural}, statistical physics~\cite{tanaka2000meanfield, Brody2008eqilibrium} and thermodynamics~\cite{uffink1999uncertainty, crooks2007measuring,ito2018infogeo, dechant2018infogeo, weinhold1975metric, ruppeiner1979thermodynamics, salamon1983thermodynamic, feng2008length, sivak2012thermodynamic, polettini2013nonconvexity, machta2015dissipation, lahiri2016universal, tajima2017efficiency, rotskoff2017geometric, takahashi2017shortcuts, shimazaki2018neurons}. The projection theorem~\cite{cover2012elements, amari2001hierarchy} plays a crucial role in applications of information geometry. For example, the projection theorem unifies the conventional definitions of information measures such as the mutual information, the transfer entropy and several measures in the integrated information theory~\cite{amari2016information, oizumi2016unified,amari2017integration}. 

The integrated information theory seeks for measures of inseparability of networks~\cite{Tononi2004information,balduzzi2008integrated,Barrett2011practical,Oizumi2014integrated,Ay2015information, Tononi2016integrated,oizumi2016unified,Oizumi2016measuring,amari2017integration,Tegmark2016improved,mediano2019beyond}. Several measures have been proposed by considering different ways of dividing networks~\cite{oizumi2016unified,amari2017integration, Tegmark2016improved}. A possible promising measure of information integration is the stochastic interaction~\cite{Barrett2011practical, Ay2015information}, that quantifies inseparability of stochastic dynamics in two interacting systems. 

In the field of stochastic thermodynamics~\cite{sekimoto2010stochastic, seifert2012stochastic}, a similar problem of inseparability takes place. For example, in the context of Maxwell's demon, information thermodynamic measures of the correlation between two interacting dynamics have been discussed~\cite{parrondo2015thermodynamics, sagawa2010generalized, still2012thermodynamics,sagawa2012fluctuation,ito2013information,hartich2014stochastic,horowitz2014thermodynamics,ito2015maxwell,spinney2016transfer,ito2016backward,crooks2016marginal,Auconi2018backward}. For two interacting dynamics, we introduce a measure of information thermodynamics, namely the partial entropy production for the subsystem~\cite{ito2013information,hartich2014stochastic,horowitz2014thermodynamics}. If two interacting dynamics are well separated, the sum of the partial entropy productions for each subsystem are equivalent to the total entropy production. This fact is known as the additivity of the entropy productions. If two interacting dynamics are not well separated, this additivity is generally violated.

In this letter, we introduce a novel framework of stochastic thermodynamics based on information geometry. We introduce several submanifolds related to backward dynamics, and the total entropy production and the partial entropy production can be considered to be given by the projections of the entire system onto these submanifolds. From the inclusion property of these submanifolds, we obtain a geometric interpretation of the additivity of the entropy productions. This interpretation clarifies a relationship between the violation of the additivity and the stochastic interaction. Additionally, our framework leads to a novel expression of the entropy production by considering an optimization problem to minimize the Kullback-Leibler divergence. We analytically illustrate our results by using the spin models. 

\textit{The projection theorem.--}
We first introduce the projection theorem in information geometry, which is a differential geometrical theory for the manifold of the probability distribution~\cite{cover2012elements, amari2001hierarchy}. In information geometry, a Riemannian metric is given by the Fisher information matrix and a dual pair of affine connections are defined~\cite{amari2007methods}.  Let $p_{\boldsymbol{S}}(\boldsymbol{s})$ be the joint probability, where ${\boldsymbol{S}}=\{S_1, ..., S_N \}$ is the set of random variables and ${\boldsymbol{s}}=\{s_1, ..., s_N \}$ is the set of events, respectively. In information geometry, the set of the joint probabilities is considered as a manifold. A subset of probabilities gives a submanifold $\mathcal{M}$, and a probability $p_{\boldsymbol{S}}(\boldsymbol{s})$ corresponds to a point.

\begin{figure}[tb]
\centering
\includegraphics[width=8cm]{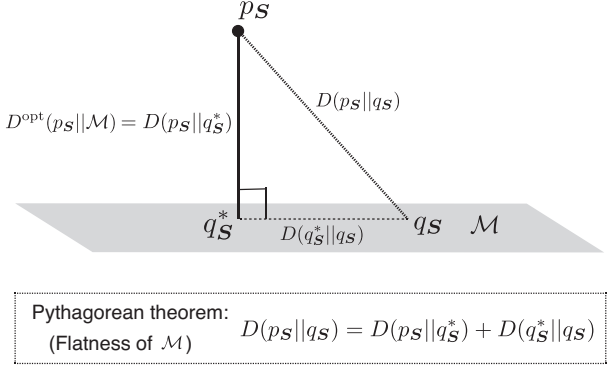}
\caption{Schematic of the projection theorem. The subset of probabilities gives a submanifold $\mathcal{M}$, and the probability $p$ corresponds to a point. If $\mathcal{M}$ is flat, we have a unique solution $q^*_{\boldsymbol{S}}$ of the optimization problem to minimize the Kullback-Leibler divergence between the probability $p$ and the probability $q_{\boldsymbol{S}} \in \mathcal{M}$. The flatness of the manifold is given by the Pythagorean theorem, and the solution $q_{\boldsymbol{S}}^*$ is the projection onto the flat submanifold $\mathcal{M}$.}
\label{fig0}
\end{figure}

We now consider an optimization problem to minimize the Kullback-Leibler divergence between two probabilities $p_{\boldsymbol{S}}(\boldsymbol{s})$ and $q_{\boldsymbol{S}}(\boldsymbol{s})$,
\begin{align} 
&D^{\rm opt} (p_{\boldsymbol{S}}|| \mathcal{M}) := {\rm min}_{q_{\boldsymbol{S}} \in \mathcal{M}} D(p_{\boldsymbol{S}}||q_{\boldsymbol{S}}), \\
&D(p_{\boldsymbol{S}}||q_{\boldsymbol{S}}) := \sum_{\boldsymbol{s}} p_{\boldsymbol{S}}(\boldsymbol{s}) \ln \frac{p_{\boldsymbol{S}}(\boldsymbol{s})}{ q_{\boldsymbol{S}}(\boldsymbol{s})},
\end{align} 
when $q_{\boldsymbol{S}} (\boldsymbol{s})$ is in a submanifold $\mathcal{M}$.
If the submanifold $\mathcal{M}$ is flat, we have the unique solution $q^*_{\boldsymbol{S}} \in \mathcal{M}$ that satisfies $D^{\rm opt} (p_{\boldsymbol{S}}|| \mathcal{M}) = D(p_{\boldsymbol{S}}||q^*_{\boldsymbol{S}})$. This unique solution $q^*_{\boldsymbol{S}}$ can be interpreted as the projection from the point $p_{\boldsymbol{S}}$ onto the flat submanifold $\mathcal{M}$. In Fig.~\ref{fig0}, we show an intuitive schematic of the projection theorem. 

This projection can be understood by considering the Pythagorean theorem 
\begin{align} 
D(p_{\boldsymbol{S}}||q_{\boldsymbol{S}})= D(p_{\boldsymbol{S}}||q^*_{\boldsymbol{S}}) +D(q^*_{\boldsymbol{S}}||q_{\boldsymbol{S}}) ,
\end{align} 
for any probability $q_{\boldsymbol{S}}$ on the flat submanifold $\mathcal{M}$~\cite{amari2007methods}. This Pythagorean theorem can be regarded as the definition of the flatness of a submanifold $\mathcal{M}$. In information geometry, the Pythagorean theorem holds when the geodesic connecting $p_{\boldsymbol{S}}$ and $q^*_{\boldsymbol{S}}$ is orthogonal to the dual geodesic connecting $q^*_{\boldsymbol{S}}$ and $q_{\boldsymbol{S}}$. From the nonnegativity of the Kullback-Leibler divergence $D(q^*_{\boldsymbol{S}}||q_{\boldsymbol{S}}) \geq 0$, we obtain the fact that $q^*_{\boldsymbol{S}}$ is the unique solution of an optimization problem
\begin{align} 
D(p_{\boldsymbol{S}}||q_{\boldsymbol{S}}) \geq D(p_{\boldsymbol{S}}||q^*_{\boldsymbol{S}}) = D^{\rm opt} (p_{\boldsymbol{S}}|| \mathcal{M}).
\end{align} 

\textit{The total entropy production and projection.--} We here consider a Markov process. Let $\boldsymbol{Z}$ and $\boldsymbol{Z}'$ be random variables of the state of a system $\mathcal{Z}$ at time $t$ and $t+dt$, respectively. Let $p_{\boldsymbol{Z}, \boldsymbol{Z'}}(\boldsymbol{z}, \boldsymbol{z'})$ be the joint probability of the states $\boldsymbol{s} =\{ \boldsymbol{z}, \boldsymbol{z'} \}$ corresponding to random variables $\boldsymbol{S}= \{ \boldsymbol{Z}, \boldsymbol{Z}' \}$. The transition probability is given by $T(\boldsymbol{z'}, \boldsymbol{z}) := p_{\boldsymbol{Z}'| \boldsymbol{Z}}(\boldsymbol{z}'| \boldsymbol{z})$,
where the conditional probability is defined as $p_{\boldsymbol{Z'}| \boldsymbol{Z}}(\boldsymbol{z'}| \boldsymbol{z}) := p_{\boldsymbol{Z'}, \boldsymbol{Z}}(\boldsymbol{z'}, \boldsymbol{z})/ p_{\boldsymbol{Z}}(\boldsymbol{z})= p_{\boldsymbol{S}}(\boldsymbol{s})/ [\sum_{\boldsymbol{z'}} p_{\boldsymbol{S}}(\boldsymbol{s})]$. Because the transition probability $T(\boldsymbol{z'}, \boldsymbol{z})$ is a function of $(\boldsymbol{z'},\boldsymbol{z})$, we can define a new quantity $T(\boldsymbol{z}, \boldsymbol{z'})$ by replacing $\boldsymbol{z}$ with $\boldsymbol{z'}$. Remark that $T(\boldsymbol{z'}, \boldsymbol{z})$ is not equal to the conditional probability $p_{\boldsymbol{Z}| \boldsymbol{Z'}}(\boldsymbol{z}| \boldsymbol{z'}) := p_{\boldsymbol{S}}(\boldsymbol{s})/ [\sum_{\boldsymbol{z}} p_{\boldsymbol{S}}(\boldsymbol{s})]$.

In stochastic thermodynamics~\cite{seifert2012stochastic}, the total entropy production $\sigma^{\mathcal{Z}}_{\rm tot}$ is defined as the sum of the entropy changes,
\begin{align}
\sigma^{\mathcal{Z}}_{\rm tot} &: =\sigma^{\mathcal{Z}}_{\rm sys}+ \sigma^{\mathcal{Z}}_{\rm bath}.
\end{align} 
The entropy change of the system $\sigma^{\mathcal{Z}}_{\rm sys}$ is defined as the Shannon entropy change from time $t$ to $t+dt$.
\begin{align}
\sigma^{\mathcal{Z}}_{\rm sys}&: =H(\boldsymbol{Z'}) - H(\boldsymbol{Z}),
\end{align} 
where  $H(\boldsymbol{Z}) = - \sum_{\boldsymbol{z}} p_{\boldsymbol{Z}} (\boldsymbol{z})\ln  p_{\boldsymbol{Z}} (\boldsymbol{z})$ is the Shannon entropy.
The entropy change of the heat bath $\sigma^{\mathcal{Z}}_{\rm bath}$ is defined as 
\begin{align}
\sigma^{\mathcal{Z}}_{\rm bath}&: = \mathbb{E} \left[ \ln \frac{T(\boldsymbol{z'}, \boldsymbol{z})}{T(\boldsymbol{z}, \boldsymbol{z'})} \right]   = \mathbb{E} \left[ -\ln  T(\boldsymbol{z}, \boldsymbol{z'}) \right]- H( \boldsymbol{Z'}| \boldsymbol{Z}) 
\end{align} 
where the symbol $\mathbb{E}[ \cdots ] := \sum_{\boldsymbol s} p_{\boldsymbol S} (\boldsymbol s) \cdots$ denotes the expected value and $H( \boldsymbol{Z'}| \boldsymbol{Z}) := H(\boldsymbol{Z'}, \boldsymbol{Z} ) - H(\boldsymbol{Z})$ is the conditional Shannon entropy. The entropy change of the heat bath can be regarded as the difference between the conditional cross entropy $ \mathbb{E} \left[ -\ln  T(\boldsymbol{z}, \boldsymbol{z'}) \right]$ and the conditional Shannon entropy. The nonnegativity of the entropy production is known as the second law of thermodynamics.
 If the entropy production is zero, the system is reversible and the detailed balance $p_{\boldsymbol{Z}}(\boldsymbol{z})T(\boldsymbol{z'},\boldsymbol{z})=p_{\boldsymbol{Z'}}(\boldsymbol{z'})T(\boldsymbol{z}, \boldsymbol{z'})$ holds~\cite{sm}. Hence, $\sigma^{\mathcal{Z}}_{\rm tot}$ quantifies irreversibility of dynamics.

\begin{figure}[tb]
\centering
\includegraphics[width=7cm]{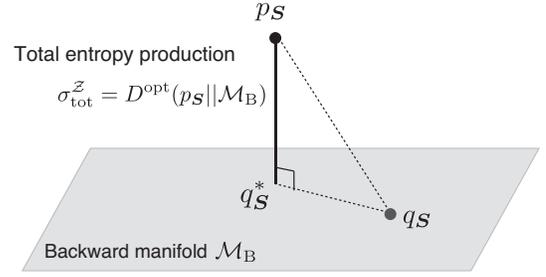}
\caption{Schematic of the total entropy production and the projection onto the backward manifold $\mathcal{M}_{\rm B}$. The entropy production $\sigma^{\mathcal{Z}}_{\rm tot}$ is given by the minimum length from the backward manifold $D^{\rm opt} (p_{\boldsymbol{S} }|| \mathcal{M}_{\rm B})$.}
\label{fig1}
\end{figure}
We show that the total entropy production can be obtained by the projection of $p_{\boldsymbol{S} }$ onto a submanifold, called 
{\it the backward manifold}. 
The backward manifold $\mathcal{M}_{\rm B}$ is defined as the set of probabilities $q_{\boldsymbol{S}}$ satisfying
\begin{align}
\mathcal{M}_{\rm B} = \{ q_{\boldsymbol{S}}  |  q_{\boldsymbol{S}}(\boldsymbol{s}) =  q_{\boldsymbol{Z'}}(\boldsymbol{z'})T(\boldsymbol{z}, \boldsymbol{z'}) \},
\end{align} 
where $q_{\boldsymbol{Z'}}(\boldsymbol{z'}) = \sum_{\boldsymbol{z}} q_{\boldsymbol{S}}(\boldsymbol{s})$ and $T(\boldsymbol{z}, \boldsymbol{z'})$ is defined from $p_{\boldsymbol{S}}  (\boldsymbol s)$. The backward manifold consists of probabilities such that backward dynamics from $\boldsymbol{Z'}$ to $\boldsymbol{Z}$ is equal to the transition probability of $p_{\boldsymbol{S}}$. The backward manifold is uniquely determined by $p_{\boldsymbol{S}}$. The total entropy production of the Markov process is given by
\begin{align}
\sigma^{\mathcal{Z}}_{\rm tot} =D^{\rm opt} (p_{\boldsymbol{S} }|| \mathcal{M}_{\rm B}),
\label{opt1}
\end{align} 
which is the first main result of this letter. This result means that the total entropy production can be regarded as the minimum length of $p_{\boldsymbol{S}}$ to the backward manifold (see also Fig.~\ref{fig1}). To prove Eq.~(\ref{opt1}), we introduce the joint probability $q^*_{\boldsymbol{S}} ({\boldsymbol{s}}) :=p_{\boldsymbol{Z'}}( \boldsymbol{z'}) T(\boldsymbol{z}, \boldsymbol{z}') \in \mathcal{M}_{\rm B}$, the entropy production is given by the Kullback-Leibler divergence $\sigma^{\mathcal{Z}}_{\rm tot} = D(p_{\boldsymbol{S}}||q^*_{\boldsymbol{S}})$~\cite{kawai2007dissipation}.
Because the following Pythagorean theorem 
\begin{align}
D(p_{\boldsymbol{S}}||q_{\boldsymbol{S}})= D(p_{\boldsymbol{S}}||q^*_{\boldsymbol{S}}) + D(q^*_{\boldsymbol{S}}||q_{\boldsymbol{S}})
\end{align} 
is valid for any $q_{\boldsymbol{S}} \in \mathcal{M}_{\rm B}$~\cite{sm}, we obtain the first main result Eq.~(\ref{opt1}). 

\textit{The second law of information thermodynamics.--} 
We next consider the situation that the system $\mathcal{Z}$ consists of two subsystems $\mathcal{X}$ and $\mathcal{Y}$, and random variables $\boldsymbol{Z}$ and $\boldsymbol{Z'}$ are given by $\boldsymbol{Z}= \{\boldsymbol{X},\boldsymbol{Y}\}$ and $\boldsymbol{Z'}= \{\boldsymbol{X'},\boldsymbol{Y'}\}$, respectively. The transition probability of the subsystem $\mathcal{X}$ for fixed states $\{ \boldsymbol{y}, \boldsymbol{y}' \}$ is given by $T^{\mathcal{X}}(\boldsymbol{z}', \boldsymbol{z}) :=p_{\boldsymbol{X}' |\boldsymbol{Y}',\boldsymbol{Z}} (\boldsymbol{x}' | \boldsymbol{y}',\boldsymbol{z})$.

The partial entropy production for the subsystem $\mathcal{X}$ is defined as
\begin{align}
&\sigma^{\mathcal{X}}_{\rm partial} := \sigma^{\mathcal{X}}_{\rm sys}+ \sigma^{\mathcal{X}}_{\rm bath} -  \Theta^{\mathcal{X} \to \mathcal{Y}}, \label{samefrom} \\
&\sigma^{\mathcal{X}}_{\rm sys}= H(\boldsymbol{X'})-H(\boldsymbol{X}),\\
&\sigma^{\mathcal{X}}_{\rm bath}= \mathbb{E} \left[ \ln \frac{T^{\mathcal{X}}(\boldsymbol{z}', \boldsymbol{z}) }{T^{\mathcal{X}}(\boldsymbol{z}, \boldsymbol{z}') } \right], \\
&\Theta^{\mathcal{X} \to \mathcal{Y}}= I(\boldsymbol{X'};\{\boldsymbol{Y},\boldsymbol{Y'}\}) -I(\boldsymbol{X};\{\boldsymbol{Y},\boldsymbol{Y'} \}).
\end{align} 
where $ I(\boldsymbol{Z};\boldsymbol{Z'}) = H(\boldsymbol{Z}) - H(\boldsymbol{Z}|\boldsymbol{Z'})$ is the mutual information between two random variables $\boldsymbol{Z}$ and $\boldsymbol{Z'}$. The additional term $\Theta^{\mathcal{X} \to \mathcal{Y}}$ quantifies dynamic information flow from the subsystem $\mathcal{X}$ to the subsystem $\mathcal{Y}$. Thus, the nonnegativity of the partial entropy production can be regarded as the second law of information thermodynamics for the subsystem $\sigma^{\mathcal{X}}_{\rm sys}+ \sigma^{\mathcal{X}}_{\rm bath}  \geq \Theta^{\mathcal{X} \to \mathcal{Y}}$, which implies a trade-off relationship between the entropy changes $\sigma^{\mathcal{X}}_{\rm sys}+ \sigma^{\mathcal{X}}_{\rm bath}$ and information flow $\Theta^{\mathcal{X} \to \mathcal{Y}}$. The partial entropy production for the subsystem $\mathcal{X}$ quantifies local irreversibility of dynamics in the system $\mathcal{X}$. The partial entropy production vanishes if dynamics in the system $\mathcal{X}$ are locally reversible, that is $p_{\boldsymbol{Z}, \boldsymbol{Y'}}(\boldsymbol{z}, \boldsymbol{y'}) T^{\mathcal{X}}(\boldsymbol{z}', \boldsymbol{z})=p_{\boldsymbol{Z'}, \boldsymbol{Y}}(\boldsymbol{z'}, \boldsymbol{y})T^{\mathcal{X}}(\boldsymbol{z}, \boldsymbol{z'})$.

We here show that the partial entropy production can also be derived from the projection of $p_{\boldsymbol{S}}$ onto {\it the local backward manifold}. The local backward manifold of the system $\mathcal{X}$ is defined as the set of probabilities such that
\begin{align}
    \mathcal{M}_{\rm LB}^{\mathcal{X}} = \left\{ q_{\boldsymbol{S}} \left|  q_{\boldsymbol{S}}(\boldsymbol{s})=q_{\boldsymbol{Y}, \boldsymbol{Z'}}(\boldsymbol{y}, \boldsymbol{z'}) T^{\mathcal{X}}(\boldsymbol{z}, \boldsymbol{z}') \right. \right\},
 \end{align}
 where $q_{\boldsymbol{Y}, \boldsymbol{Z'}}(\boldsymbol{y}, \boldsymbol{z'})= \sum_{\boldsymbol{x}} q_{\boldsymbol{S}}(\boldsymbol{s})$ and $T^{\mathcal{X}}(\boldsymbol{z}, \boldsymbol{z'})$ is defined from $p_{\boldsymbol{S}}  (\boldsymbol s)$. The local backward manifold means the set of probabilities such that local backward dynamics from $\boldsymbol{X'}$ to $\boldsymbol{X}$ is equal to the transition probability in $\mathcal{X}$ of $p_{\boldsymbol{S} }$.  The partial entropy production of the subsystem $\mathcal{X}$ is given by
\begin{align}
\sigma^{\mathcal{X}}_{\rm partial} =D^{\rm opt} (p_{\boldsymbol{S} }|| \mathcal{M}_{\rm LB}^{\mathcal{X}}),
\label{opt2}
\end{align} 
which is the second main result of this letter. To prove Eq.~(\ref{opt2}), we introduce the probability $q^{\mathcal{X}*}_{\boldsymbol{S}}(\boldsymbol{s})=T^{\mathcal{X}}(\boldsymbol{z}, \boldsymbol{z}')  p_{\boldsymbol{Y}, \boldsymbol{Z'}}(\boldsymbol{y}, \boldsymbol{z'}) \in \mathcal{M}_{\rm LB}^{\mathcal{X}}$. Because we can show the following expression
\begin{align}
\sigma^{\mathcal{X}}_{\rm partial} = D(p_{\boldsymbol{S} }||q^{\mathcal{X}*}_{\boldsymbol{S}}),
\end{align}
and the Pythagorean theorem
\begin{align}
D(p_{\boldsymbol{S}}||q^{\mathcal{X}}_{\boldsymbol{S}}) &= D(p_{\boldsymbol{S}}||q^{\mathcal{X}*}_{\boldsymbol{S}}) + D(q^{\mathcal{X}*}_{\boldsymbol{S}}||q^{\mathcal{X}}_{\boldsymbol{S}}),
\label{Pythagorean}
\end{align} 
for any $q^{\mathcal{X}}_{\boldsymbol{S}} \in  \mathcal{M}_{\rm LB}^{\mathcal{X}}$, we obtain the second main result Eq.~(\ref{opt2}). If we introduce the quantities for the subsystem $\mathcal{Y}$ such as $(T^{\mathcal{Y}}, \sigma^{\mathcal{Y}}_{\rm partial}, \sigma^{\mathcal{Y}}_{\rm sys}, \sigma^{\mathcal{Y}}_{\rm bath}, \Theta^{\mathcal{Y} \to \mathcal{X}}, \mathcal{M}_{\rm LB}^{\mathcal{Y}})$ by replacing $(\boldsymbol{X}, \boldsymbol{X'})$ with $(\boldsymbol{Y}, \boldsymbol{Y'})$, we obtain the same results Eqs.~(\ref{samefrom})-(\ref{Pythagorean}) for the subsystem $\mathcal{Y}$.

\begin{figure}[tb]
\centering
\includegraphics[width=8cm]{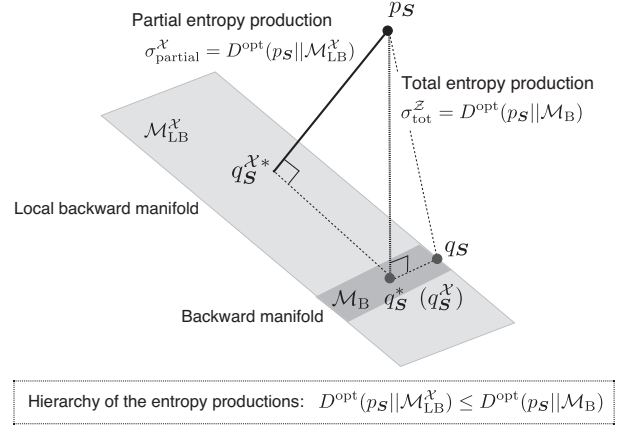}
\caption{Schematic of the partial entropy production and the hierarchy of the entropy productions. Because the local backward manifold includes the backward manifold, the partial entropy production is always smaller than the total entropy production.}
\label{fig2}
\end{figure}
We notify that our geometric interpretation provides the hierarchy of the entropy productions. Because the backward manifold is a submanifold of the local backward manifold $\mathcal{M}_{\rm B} \subset \mathcal{M}_{\rm LB}^{\mathcal{X}}$, we obtain the hierarchy $D^{\rm opt} (p_{\boldsymbol{S} }|| \mathcal{M}_{\rm LB}^{\mathcal{X}}) \leq  D^{\rm opt} (p_{\boldsymbol{S} }|| \mathcal{M}_{\rm B})$,
or equivalently 
\begin{align}
\sigma^{\mathcal{X}}_{\rm partial} \leq \sigma^{\mathcal{Z}}_{\rm tot}.
\label{hierarchy}
\end{align}
This hierarchy of the entropy productions implies that the second law of information thermodynamics always gives a tighter bound than the second law of thermodynamics (see also Fig.3). Moreover, if the subsystem $\mathcal{X}_1$ includes the subsystem $\mathcal{X}_2$, we obtain the hierarchy of the entropy productions 
\begin{align}
\sigma^{\mathcal{X}_{2}}_{\rm partial} \leq \sigma^{\mathcal{X}_{1}}_{\rm partial},
\end{align} 
from the inclusion property $\mathcal{M}_{\rm LB}^{\mathcal{X}_{1}} \subset \mathcal{M}_{\rm LB}^{\mathcal{X}_{2}}$.
This hierarchy clarifies the relationships between the second laws of information thermodynamics in complex systems.  

{\it The stochastic interaction.--} We here introduce the stochastic interaction~\cite{Barrett2011practical,Ay2015information} as a measure of bidirectional information flow. The stochastic interaction~\cite{Barrett2011practical,Ay2015information} is defined as 
\begin{align}
&\Phi_{\rm SI} := D(p_{\boldsymbol{Z}, \boldsymbol{Z'}}||p_{\boldsymbol{X'}| \boldsymbol{Z}} p_{\boldsymbol{Y}'| \boldsymbol{Z}}p_{\boldsymbol{Z}} ),
\end{align}
This quantity is zero if the stochastic process satisfies the bipartite condition $\mathcal{C}_{\rm BI}: p_{\boldsymbol{Z'}|\boldsymbol{Z}}(\boldsymbol{z'}|\boldsymbol{z}) = p_{\boldsymbol{X'}|\boldsymbol{Z}}(\boldsymbol{x'}|\boldsymbol{z})p_{\boldsymbol{Y'}|\boldsymbol{Z}}(\boldsymbol{y'}|\boldsymbol{z})$. The bipartite condition $\mathcal{C}_{\rm BI}$ means that two transitions in $\mathcal{X}$ and $\mathcal{Y}$ are statistically independent, because the transition probability $T^{\mathcal{X}}(\boldsymbol{z}', \boldsymbol{z})=  p_{\boldsymbol{X'}|\boldsymbol{Z}}(\boldsymbol{x'}|\boldsymbol{z})$ does not depend on $\boldsymbol{y}'$ under the bipartite condition.  We also define the stochastic interaction for backward dynamics as
\begin{align}
&\Phi_{\rm SI}^{\dagger} := D(p_{\boldsymbol{Z}, \boldsymbol{Z'}}||p_{\boldsymbol{X}| \boldsymbol{Z'}} p_{\boldsymbol{Y}| \boldsymbol{Z'}}p_{\boldsymbol{Z'}} ),
\end{align}
which exactly vanishes under the backward bipartite condition
$\mathcal{C}_{\rm BI}^{*} : p_{\boldsymbol{Z}| \boldsymbol{Z'}}(\boldsymbol{z}| \boldsymbol{z'}) =p_{\boldsymbol{X}| \boldsymbol{Z'}}(\boldsymbol{x}|\boldsymbol{z'}) p_{\boldsymbol{Y}| \boldsymbol{Z'}}(\boldsymbol{y}|\boldsymbol{z'})$.

While the stochastic interactions are measures of bidirectional information flow, the dynamic information flow $\Theta^{\mathcal{X} \to \mathcal{Y}}$ is a measure of directed information flow. $\Theta^{\mathcal{X} \to \mathcal{Y}}$ can be decomposed into the mutual information difference $\Delta \mathcal{I}$ and the measures of directed information flow, i.e., the transfer entropy $I(\boldsymbol{X};\boldsymbol{Y'}|\boldsymbol{Y})$~\cite{schreiber2000transfer, barnett2014transfer} and the backward transfer entropy $ I(\boldsymbol{X'};\boldsymbol{Y}|\boldsymbol{Y'})$~\cite{ito2016backward},
\begin{align}
\Theta^{\mathcal{X} \to \mathcal{Y}} &=\Delta \mathcal{I} + I(\boldsymbol{X'};\boldsymbol{Y}|\boldsymbol{Y'}) -I(\boldsymbol{X};\boldsymbol{Y'}|\boldsymbol{Y}), \\
\Delta \mathcal{I} &:= I(\boldsymbol{X'};\boldsymbol{Y'}) -I(\boldsymbol{X};\boldsymbol{Y}),
\end{align} 
where $I(\boldsymbol{Z};\boldsymbol{Z'} | \boldsymbol{Z''}) := H(\boldsymbol{Z}| \boldsymbol{Z''})  -H(\boldsymbol{Z}|\boldsymbol{Z'}, \boldsymbol{Z''}) $ is the conditional mutual information between $\boldsymbol{Z}$ and $\boldsymbol{Z'}$ under the condition $\boldsymbol{Z''}$. To compare the dynamic information flow with the stochastic interaction, we consider the bidirectional information flow by considering the sum of $\Theta^{\mathcal{X} \to \mathcal{Y}}$ and $\Theta^{\mathcal{Y} \to \mathcal{X}}$. The relationship between the stochastic interaction and the dynamic information flow is given by
\begin{align}
\Theta^{\mathcal{X} \to \mathcal{Y}} +\Theta^{\mathcal{Y} \to \mathcal{X}} - \Delta \mathcal{I} &= \Phi_{\rm SI} -\Phi_{\rm SI}^{\dagger}.
\end{align} 

{\it Additivity and information integration.--} 
We next discuss the additivity of the partial entropy productions. We show that the violation of the additivity is related to a measure of integrated information, i.e., stochastic interaction. Under the bipartite condition $\mathcal{C}_{\rm BI}$, we have the additivity of the entropy productions up to the order $\mathcal{O}(dt^2)$~\cite{hartich2014stochastic}, 
\begin{align}
\sigma_{\rm tot}^{\mathcal{Z}} =\sigma_{\rm partial}^{\mathcal{X}} + \sigma_{\rm partial}^{\mathcal{Y}}.
\label{add}
\end{align}
From Eq.~(\ref{add}), the hierarchy Eq.~(\ref{hierarchy}) is equivalent to the second law of information thermodynamics for the subsystem $\mathcal{Y}$, that is $\sigma_{\rm partial}^{\mathcal{Y}} \geq 0$.
If time evolution of two systems are strongly correlated, the assumption of the bipartite condition is not valid, and the additivity Eq.~(\ref{add}) is violated. The amount of the violation is given by the stochastic interactions and the additional term
\begin{align}
&\sigma_{\rm tot}^{\mathcal{Z}} -\sigma_{\rm partial}^{\mathcal{X}} - \sigma_{\rm partial}^{\mathcal{Y}} =\Phi_{\rm bath} +\Phi_{\rm SI} - \Phi_{\rm SI}^{\dagger}, \\
&\Phi_{\rm bath} := \sigma^{\mathcal{Z}}_{\rm bath} - \sigma^{\mathcal{X}}_{\rm bath}  - \sigma^{\mathcal{Y}}_{\rm bath}.
\label{additivity}
\end{align}
The additional term $\Phi_{\rm bath}$ quantifies to what extent the additivity is violated in the heat bathes. This measure $\Phi_{\rm bath}$ can be considered as a novel measure of information integration for thermal systems, because the entropy change does not attract much attention in integrated information theory.

\begin{figure}[tb]
\centering
\includegraphics[width=7cm]{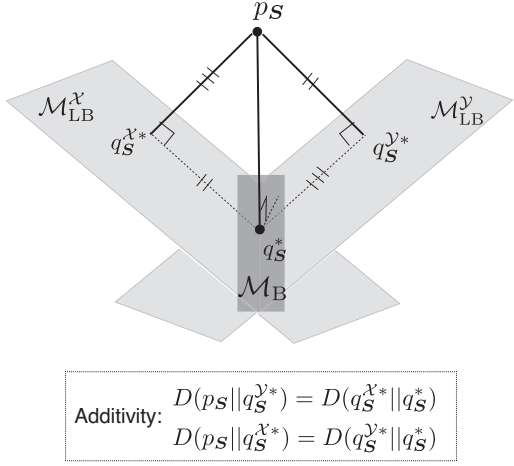}
\caption{Schematic of the additivity and the rectangle. Under the both bipartite conditions $\mathcal{C}_{\rm BI}$ and $\mathcal{C}_{\rm BI}^{*}$, the backward manifold is equal to the intersection of the local backward manifolds. The additivity of the entropy production indicates that the parallel sides of a quadrangle $(p_{\boldsymbol S},{q}^{\mathcal{X}*}_{\boldsymbol S}, {q}^{*}_{\boldsymbol S}, {q}^{\mathcal{Y}*}_{\boldsymbol S})$ have the same length.}
\label{fig3}
\end{figure}

We show a geometrical condition of this additivity under the both bipartite conditions $\mathcal{C}_{\rm BI}$ and $\mathcal{C}_{\rm BI}^{*}$. The both bipartite conditions implies the relationship between three manifolds
\begin{align}
\mathcal{M}_{\rm B} = \mathcal{M}^{\mathcal{X}}_{\rm LB} \cap \mathcal{M}^{\mathcal{Y}}_{\rm LB}. 
\end{align} 
Because Eq.~(\ref{add}) can be written as 
\begin{align}
D(p_{\boldsymbol{S}}||q^{*}_{\boldsymbol{S}})=D(p_{\boldsymbol{S}}|| q^{\mathcal{X}*}_{\boldsymbol{S}})+D(p_{\boldsymbol{S}}|| q^{\mathcal{Y}*}_{\boldsymbol{S}}),
\end{align}
we obtain the following relationship
\begin{align}
D(p_{\boldsymbol{S}}|| q^{\mathcal{X}*}_{\boldsymbol{S}}) &=D(q^{\mathcal{Y}*}_{\boldsymbol{S}}|| q^{*}_{\boldsymbol{S}}),\label{additivity2} \\
D(p_{\boldsymbol{S}}|| q^{\mathcal{Y}*}_{\boldsymbol{S}}) &=D(q^{\mathcal{X}*}_{\boldsymbol{S}}|| q^{*}_{\boldsymbol{S}}),
\label{additivity2-1}
\end{align}
from the Pythagorean theorem Eq.~(\ref{Pythagorean}). The equations~(\ref{additivity2}) and (\ref{additivity2-1}) implies that the parallel sides of a quadrangle have the same length. Therefore, the additivity Eq.~(\ref{add}) can be understood from {\it the rectangle condition} in information geometry (Fig.~\ref{fig3}). The measures of information integration $\Phi_{\rm bath} +\Phi_{\rm SI} - \Phi_{\rm SI}^{\dagger}$ quantifies a distortion of this rectangle.

{\it Example I: Single spin model.--} We illustrate the main result Eq.~(\ref{opt1}) by the single spin model~\cite{sm}. Let ${\boldsymbol Z}=\{S_1\}$ and ${\boldsymbol Z'}= \{S_2\}$ be random variables of the spin at time $t$ and $t+dt$, respectively. The each spin has the binary state $s_i \in \{0, 1 \}$. The joint probability is generally given by the exponential family
\begin{align}  
p_{{\boldsymbol S}}^{\hat{\boldsymbol{\theta}}} (s_1, s_2) =& \exp \left[ \sum_{i} s_i \hat{\theta}^i +\sum_{i<j} s_i s_j \hat{\theta}^{ij} - \phi_{{\boldsymbol S}}  (\boldsymbol{\hat{\theta}}) \right],
\end{align} 
where $\hat{\boldsymbol{\theta}} = \{ \hat{\theta}^1, \hat{\theta}^2, \hat{\theta}^{12}\}$ is the set of parameters, and $\phi_{{\boldsymbol S}}(\boldsymbol{\hat{\theta}})$ is the normalization factor that satisfies $\sum_{{\boldsymbol s}} p_{{\boldsymbol S}}^{\hat{\boldsymbol{\theta}}}(\boldsymbol{s}) =1$. The number of the elements in ${\hat{\boldsymbol{\theta}}}$ is $(2^2 -1) =3$, so the set of the probabilities $p_{{\boldsymbol S}}^{\hat{\boldsymbol{\theta}}}$ can be represented by $3$-dimensional submanifold. The backward manifold $\mathcal{M}_{\rm B}$ is given by the constraint of the parameters
\begin{align}  
\mathcal{M}_{\rm B}  =  \{ p_{{\boldsymbol S}}^{{\boldsymbol{\theta}}} | \theta^1 = \hat{\theta^2},  \theta^{12} = \hat{\theta^{12}}\}. 
\end{align} 
Because a free parameter is $\theta^2$, the backward manifold for the single spin model is $1$-dimensional. 

Our result Eq.~(\ref{opt1}) can be rewritten as the optimization problem of $\theta^2$,
 \begin{align}
\sigma^{\mathcal{Z}}_{\rm tot} =& {\rm min}_{\theta^2}  \left. D(p_{{\boldsymbol S}}^{\hat{\boldsymbol{\theta}}}|| p_{{\boldsymbol S}}^{{\boldsymbol{\theta}}} ) \right|_{\theta^1 = \hat{\theta^2},  \theta^{12} = \hat{\theta^{12}}}\\
=& \mathbb{E} [s_1] (\hat{\theta}^1 - \hat{\theta}^2) - \phi_{{\boldsymbol S}}  (\hat{\theta}^1 , \hat{\theta}^2, \hat{\theta}^{12}) \nonumber\\
&+  {\rm min}_{\theta^2}  \left[ \mathbb{E} [s_2] (\hat{\theta}^2 - \theta^2) +   \phi_{{\boldsymbol S}}  (\hat{\theta}^2 , {\theta}^2, \hat{\theta}^{12}) \right].
\end{align}
This problem can be numerically solved by using a conventional optimization tool.

{\it Example II: Two spins model.--} We next illustrate our results by the two spins model~\cite{sm}. Let ${\boldsymbol Z}=\{S_1, S_2\}$ and ${\boldsymbol Z'}= \{S_3, S_4\}$ be random variables of two spins at time $t$ and $t+dt$, respectively. The spin has the binary state $s_i \in \{0, 1 \}$. We assume the situation that the both bipartite conditions $\mathcal{C}_{\rm BI}$ and $\mathcal{C}_{\rm BI}^{*}$ holds. Under the bipartite conditions, the joint probability of the spin state is generally given by the exponential family
\begin{align}  
 p_{{\boldsymbol S}}^{\hat{\boldsymbol{\theta}}}(\boldsymbol{s})=& \exp \left[ \sum_{i} s_i \hat{\theta}^i +s_1 s_3 \hat{\theta}^{13} +s_1 s_4 \hat{\theta}^{14} \right. \nonumber\\ 
 &\left.  +s_2 s_3 \hat{\theta}^{23} +s_2 s_4 \hat{\theta}^{24} - \phi_{{\boldsymbol S}}  (\boldsymbol{\hat{\theta}}) \right].
\end{align} 

The backward manifold is given by the constraint of the parameters
\begin{align}
&\mathcal{M}_{\rm B} = \{ p_{{\boldsymbol S}}^{{\boldsymbol{\theta}}} |{\boldsymbol{\theta}}^{\mathcal X}  = \hat{\boldsymbol{\theta}}^{\mathcal X},  {\boldsymbol{\theta}}^{\mathcal Y}  = \hat{\boldsymbol{\theta}}^{\mathcal Y}\}, \\
&{\boldsymbol{\theta}}^{\mathcal X} = ({\theta}^1, {\theta}^{13}, {\theta}^{14} ), \: \: \: \: \hat{\boldsymbol{\theta}}^{\mathcal X} = (\hat{\theta}^3, \hat{\theta}^{13}, \hat{\theta}^{23} ), \\
&{\boldsymbol{\theta}}^{\mathcal Y} = ({\theta}^2, {\theta}^{24}, {\theta}^{23} ),  \: \: \: \: \hat{\boldsymbol{\theta}}^{\mathcal Y} = (\hat{\theta}^4, \hat{\theta}^{24}, \hat{\theta}^{14} ), 
 \label{mani1}
\end{align} 
where a coordinate ${\boldsymbol{\theta}}$ represents a probability on the backward manifold. Because free parameters are $\{ \theta^3, \theta^4 \}$, the backward manifold for the two spin models is $2$-dimensional.  The condition of the local backward manifolds are also given by the linear constraint of ${\boldsymbol{\theta}}$,
\begin{align}
\mathcal{M}_{\rm LB}^{\mathcal X} = \{ p_{{\boldsymbol S}}^{{\boldsymbol{\theta}}}  |{\boldsymbol{\theta}}^{\mathcal X}  = \hat{\boldsymbol{\theta}}^{\mathcal X}\}, \: \: \:
\mathcal{M}_{\rm LB}^{\mathcal Y} = \{ p_{{\boldsymbol S}}^{{\boldsymbol{\theta}}}  | {\boldsymbol{\theta}}^{\mathcal Y}  = \hat{\boldsymbol{\theta}}^{\mathcal Y}\}.
\label{mani2}
\end{align} 
Because free parameters are $\{ \theta^3, \theta^4, {\boldsymbol{\theta}}^{\mathcal Y} \}$ ($\{ \theta^3, \theta^4 ,{\boldsymbol{\theta}}^{\mathcal X} \}$), the local backward manifold  $\mathcal{M}_{\rm LB}^{\mathcal X}$ ($\mathcal{M}_{\rm LB}^{\mathcal X}$) is $5$-dimensional. The intersection of these two local backward manifolds is the backward manifold $\mathcal{M}_{\rm B} = \mathcal{M}^{\mathcal{X}}_{\rm LB} \cap \mathcal{M}^{\mathcal{Y}}_{\rm LB}$. As discussed in Example I, the total entropy production and the partial entropy productions are obtained from the optimization problems
 \begin{align}
\sigma^{\mathcal{Z}}_{\rm tot} =& {\rm min}_{\theta^3, \theta^4}  \left. D(p_{{\boldsymbol S}}^{\hat{\boldsymbol{\theta}}}|| p_{{\boldsymbol S}}^{{\boldsymbol{\theta}}} ) \right|_{ {\boldsymbol{\theta}}^{\mathcal X}  = \hat{\boldsymbol{\theta}}^{\mathcal X},  {\boldsymbol{\theta}}^{\mathcal Y}  = \hat{\boldsymbol{\theta}}^{\mathcal Y}},\\
\sigma^{\mathcal{X}}_{\rm partial} =& {\rm min}_{\theta^3, \theta^4, {\boldsymbol{\theta}}^{\mathcal Y}}  \left. D(p_{{\boldsymbol S}}^{\hat{\boldsymbol{\theta}}}|| p_{{\boldsymbol S}}^{{\boldsymbol{\theta}}} ) \right|_{  {\boldsymbol{\theta}}^{\mathcal X}  = \hat{\boldsymbol{\theta}}^{\mathcal X} },\\ 
\sigma^{\mathcal{Y}}_{\rm partial} =& {\rm min}_{\theta^3, \theta^4, {\boldsymbol{\theta}}^{\mathcal X}}  \left. D(p_{{\boldsymbol S}}^{\hat{\boldsymbol{\theta}}}|| p_{{\boldsymbol S}}^{{\boldsymbol{\theta}}} ) \right|_{ {\boldsymbol{\theta}}^{\mathcal Y}  = \hat{\boldsymbol{\theta}}^{\mathcal Y} }.
\end{align}

Without the bipartite condition $\mathcal{C}_{\rm BI}$ and $\mathcal{C}_{\rm BI}^{*}$, the joint probability is generally given by
\begin{align}  
 p_{\boldsymbol{S}}^{\hat{\boldsymbol{\theta}}}(\boldsymbol{s})=& \exp \left[ \sum_{i} s_i \hat{\theta}^i +\sum_{i<j} s_i s_j \hat{\theta}^{ij} + \sum_{i<j<k} s_i s_j s_k \hat{\theta}^{ijk}   \right. \nonumber\\ 
  &\left. +\sum_{i<j<k<l} s_i s_j s_ks_l \hat{\theta}^{ijkl}  - \phi_{\boldsymbol{S}} (\boldsymbol{\hat{\theta}}) \right].
\end{align} 
If the vector $(\hat{\theta}^{12}, \hat{\theta}^{34}, \hat{\theta}^{123}, \hat{\theta}^{134},\hat{\theta}^{124},\hat{\theta}^{234},\hat{\theta}^{1234})$ is non-zero, the bipartite conditions are violated and measures of information integration $\Phi_{\rm SI}$, $\Phi_{\rm SI}^{\dagger}$ and $\Phi_{\rm bath}$ have nonzero values.

\textit{Conclusion and discussion.--}By applying the information-geometric framework, we show the relationship between the entropy production and the stochastic interaction. Our result can be a foundation of the integrated information theory based on the physical law. We may discuss a thermodynamic cost of the information integration based on this framework.

Because the second law of information thermodynamics is essential for biochemical information processing~\cite{ito2015maxwell, barato2014efficiency, sartori2014thermodynamic, bo2015thermodynamic, ouldridge2017thermodynamics, mcgrath2017biochemical, Matsumoto2018implication}, this work would give a geometric insight into biochemical information processing. This work provides a physical validity of the integrated information theory~\cite{Oizumi2014integrated,Tononi2016integrated,oizumi2016unified, amari2017integration} for the biochemical information processing.

From a view point of thermodynamics, our results are complementary to other geometric expressions of the second law, such as the principle of Carath\`{e}odory~\cite{Caratheodory1976principle} and the maximum entropy thermodynamics~\cite{jaynes1957info, jaynes1957info2}. Our framework would be applicable to other generalizations of the entropy production, for example, thermodynamics under feedback control by selecting the backward manifolds for the feedback control~\cite{sm}.

\section{acknowledgement}
We are grateful to Hideaki Shimazaki for critical reading of the old version of this manuscript. We are also grateful to Andreas Dechant for the discussion of information geometry and the thermodynamic uncertainty. We thank Kunihiko Kaneko, Takahiro Sagawa and Tetsuhiro Hatakeyama for valuable comments. Sosuke Ito is supported by JSPS KAKENHI Grant No. JP16K17780, JP19H05796 and JST Presto Grant No. JP18070368, Japan.

\widetext

\section{Supplementary information}
\subsection{I. Review of the second law of thermodynamics in stochastic thermodynamics} 
We here review the second law of thermodynamics in stochastic thermodynamics. We start with the master equation
\begin{align}
    \frac{d}{dt} p (\boldsymbol{z}';t) =  \sum_{\boldsymbol{z}} \left[  W(\boldsymbol{z} \to \boldsymbol{z}';t)  p(\boldsymbol{z};t) -   W(\boldsymbol{z}' \to \boldsymbol{z};t) p(\boldsymbol{z}';t) \right],
    \label{mastereqsup}
\end{align}
where $p(\boldsymbol{z};t)$ is the probability of the state $\boldsymbol{z}$ at time $t$, and $W(\boldsymbol{z} \to \boldsymbol{z}';t)$ is the transition rate from the state $\boldsymbol{z}$ to the state $\boldsymbol{z}'$ at time $t$. In the notation of this paper, the probability of $\boldsymbol{z}$ is given by $p_{\boldsymbol{Z}}(\boldsymbol{z}) =p (\boldsymbol{z};t)$. From the master equation (\ref{mastereqsup}), we obtain the probability at time $t+dt$,
\begin{align}
    p(\boldsymbol{z}';t+dt) =  \sum_{\boldsymbol{z}} \left[  W(\boldsymbol{z} \to \boldsymbol{z}';t)  p(\boldsymbol{z};t) dt +(1-  W(\boldsymbol{z}' \to \boldsymbol{z};t)  dt)p(\boldsymbol{z}';t) \right].
    \label{mastereqsup2}
\end{align}
In the notation of the main text, $p_{\boldsymbol{Z}}(\boldsymbol{z})$ and $p_{\boldsymbol{Z}'}(\boldsymbol{z}')$ are given by $p_{\boldsymbol{Z}}(\boldsymbol{z}) = p({\boldsymbol{z}};t)$ and  $p_{\boldsymbol{Z}'}(\boldsymbol{z}') = p({\boldsymbol{z}'};t+dt)$, respectively. We also obtain the relationship between $p_{\boldsymbol{Z}}$ and $p_{\boldsymbol{Z}'}$ as
\begin{align}
     p_{\boldsymbol{Z}'}(\boldsymbol{z}') = p(\boldsymbol{z}';t) +\mathcal{O}(dt) =p_{\boldsymbol{Z}}(\boldsymbol{z}')+\mathcal{O}(dt).
\end{align}
The transition probability $T(\boldsymbol{z}', \boldsymbol{z})$ is given by
\begin{align}
   T(\boldsymbol{z}', \boldsymbol{z}) =
  \begin{cases}
   W(\boldsymbol{z} \to \boldsymbol{z}';t) dt & (\boldsymbol{z} \neq \boldsymbol{z}'),\\
   (1- \sum_{\boldsymbol{z} \neq \boldsymbol{z}'} W(\boldsymbol{z}' \to \boldsymbol{z};t)  dt) & (\boldsymbol{z} = \boldsymbol{z}').
  \end{cases}\label{transitionsup}
\end{align}

Here, we consider the detailed balance. The condition of the detailed balance is given by 
\begin{align}
     W(\boldsymbol{z} \to \boldsymbol{z}';t)  p(\boldsymbol{z};t) =   W(\boldsymbol{z}' \to \boldsymbol{z};t) p(\boldsymbol{z}';t)
     \label{detailedbalance}
\end{align}
for any $\boldsymbol{z}$ and $\boldsymbol{z}'$. This condition is valid if the system is in equilibrium.
By using the transition probability Eq.~(\ref{transitionsup}), we obtain another expression of the detailed balance condition Eq.~(\ref{detailedbalance}) as
\begin{align}
     T(\boldsymbol{z}', \boldsymbol{z})p_{\boldsymbol{Z}}(\boldsymbol{z})  = T(\boldsymbol{z}, \boldsymbol{z}') p_{\boldsymbol{Z}'}(\boldsymbol{z}'),
    \label{reversiblitysup}
\end{align}
where we used $W(\boldsymbol{z}' \to \boldsymbol{z};t) p(\boldsymbol{z}';t) dt = T(\boldsymbol{z},\boldsymbol{z}') p_{\boldsymbol{Z}}(\boldsymbol{z}') = T(\boldsymbol{z},\boldsymbol{z}') p_{\boldsymbol{Z}'}(\boldsymbol{z}') +\mathcal{O}(dt^2)$. Therefore, the detailed balance condition Eq.~(\ref{detailedbalance}) implies the reversibility of dynamics in the transition from $t$ to $t+dt$.
From the identity by the Bayes' rule
\begin{align}
 p_{\boldsymbol{Z}|\boldsymbol{Z}'}(\boldsymbol{z}|\boldsymbol{z}') =T(\boldsymbol{z}',\boldsymbol{z}) \frac{ p_{\boldsymbol{Z}}(\boldsymbol{z}) }{p_{\boldsymbol{Z}'}(\boldsymbol{z}')},
\end{align}
the detailed balance condition Eq.~(\ref{detailedbalance}) can be rewritten as
\begin{align}
T(\boldsymbol{z},\boldsymbol{z}') =p_{\boldsymbol{Z}|\boldsymbol{Z}'}(\boldsymbol{z}|\boldsymbol{z}').
\end{align}

Next, we discuss the second law of thermodynamics. For the master equation, the total entropy production ratio $\sigma^{\mathcal{Z}}_{\rm tot}/dt$ is defined as
\begin{align}
   \frac{\sigma^{\mathcal{Z}}_{\rm tot}}{dt} &= \sum_{\boldsymbol{z}, \boldsymbol{z}'} W(\boldsymbol{z} \to \boldsymbol{z}';t)  p(\boldsymbol{z};t) \ln \frac{W(\boldsymbol{z} \to \boldsymbol{z}';t)  p(\boldsymbol{z};t)}{W(\boldsymbol{z}' \to \boldsymbol{z};t)  p(\boldsymbol{z}';t)}.
\end{align}
If the detailed balance condition is valid, the entropy production vanishes $\sigma^{\mathcal{Z}}_{\rm tot}= 0$. By using the transition probability $T(\boldsymbol{z}'|\boldsymbol{z})$, we obtain another expression of the total entropy production
\begin{align}
   \sigma^{\mathcal{Z}}_{\rm tot}
   &= \sum_{\boldsymbol{z}, \boldsymbol{z}'|\boldsymbol{z} \neq  \boldsymbol{z}' } W(\boldsymbol{z} \to \boldsymbol{z}';t) dt p(\boldsymbol{z};t)\ln \frac{W(\boldsymbol{z} \to \boldsymbol{z}';t)  dt p(\boldsymbol{z};t)}{W(\boldsymbol{z}' \to \boldsymbol{z};t)dt  p(\boldsymbol{z}';t)}\\
    &=  \sum_{\boldsymbol{z}, \boldsymbol{z}'| \boldsymbol{z} \neq \boldsymbol{z}'} T(\boldsymbol{z}',\boldsymbol{z})  p_{\boldsymbol{Z}}(\boldsymbol{z}) \ln \frac{T(\boldsymbol{z}',\boldsymbol{z}) p_{\boldsymbol{Z}}(\boldsymbol{z})}{T(\boldsymbol{z},\boldsymbol{z}') p_{\boldsymbol{Z}'}(\boldsymbol{z}')} +\mathcal{O}(dt^2) \\
   &= \sum_{\boldsymbol{z}, \boldsymbol{z}'} T(\boldsymbol{z}',\boldsymbol{z})  p_{\boldsymbol{Z}}(\boldsymbol{z}) \ln \frac{T(\boldsymbol{z}',\boldsymbol{z}) p_{\boldsymbol{Z}}(\boldsymbol{z})}{T(\boldsymbol{z},\boldsymbol{z}') p_{\boldsymbol{Z}'}(\boldsymbol{z}')}.
  \label{suptotent}
\end{align}
To introduce two probabilities $p_{\boldsymbol{S}}(\boldsymbol{s})= T(\boldsymbol{z}',\boldsymbol{z})  p_{\boldsymbol{Z}}(\boldsymbol{z})$ and $q^*_{\boldsymbol{S}}(\boldsymbol{s})= T(\boldsymbol{z},\boldsymbol{z}') p_{\boldsymbol{Z}'}(\boldsymbol{z}')$ with $\boldsymbol{S} = \{\boldsymbol{Z},\boldsymbol{Z'} \}$ and  $\boldsymbol{s} = \{\boldsymbol{z},\boldsymbol{z'} \}$, this expression of the total entropy production Eq.~(\ref{suptotent}) can be regarded as the Kullback-Leibler divergence between two probabilities
\begin{align}
   \sigma^{\mathcal{Z}}_{\rm tot} &= \sum_{\boldsymbol{s}} p_{\boldsymbol{S}}(\boldsymbol{s})  \ln \frac{p_{\boldsymbol{S}}(\boldsymbol{s}) }{q^*_{\boldsymbol{S}}(\boldsymbol{s}) } \\
   &= D(p_{\boldsymbol{S}}||q^*_{\boldsymbol{S}}).
\end{align}

\subsection{II. The detailed calculation of Example I: Single spin model}
We here show a detailed calculation of the single spin model. The spin state at time $t$ is $\boldsymbol{z} =s_1 \in \{0,1 \}$ and the spin state at time $t+dt$ is $\boldsymbol{z}' =s_2 \in \{0,1 \}$, respectively. We here start with the master equation 
\begin{align}
    \frac{d}{dt} p (s';t) =  \sum_{s} \left[  W(s \to s';t)  p(s;t) -   W(s' \to s;t) p(s';t) \right],
\end{align}
where $p (s;t)$ is the probability of the state $s$ at time $t$ and $W(s \to s';t)$ is the transition rate from $s$ to $s'$ at time $t$. The transition probability $T(s_2,s_1)$ is given by
\begin{align}
   T(s_2,s_1) =
  \begin{cases}
   (1-  W(0 \to 1;t)  dt) & (s_1=0, s_2=0), \\
   W(0 \to 1;t) dt & (s_1=0, s_2=1),\\
   W(1 \to 0;t) dt & (s_1=1, s_2=0), \\
   (1-  W(1 \to 0;t)  dt) & (s_1=1, s_2=1).
  \end{cases}
\end{align}
The joint probability $p_{\boldsymbol{S}}(\boldsymbol{s})$ is given by
\begin{align}
   p_{\boldsymbol{S}}(\boldsymbol{s})= T(s_2,s_1)p (s_1;t) =
  \begin{cases}
   (1-  W(0 \to 1;t)  dt) p (0;t)& (s_1=0, s_2=0), \\
   W(0 \to 1;t) dtp (0;t) & (s_1=0, s_2=1),\\
   W(1 \to 0;t) dt (1-p (0;t))& (s_1=1, s_2=0),\\
   (1-  W(1 \to 0;t)  dt) (1-p (0;t))& (s_1=1, s_2=1).
  \end{cases}
  \label{supspin}
\end{align}
Here we introduce the joint probability $p^{\boldsymbol{\hat{\theta}}}_{\boldsymbol{S}}(\boldsymbol{s})$ as the exponential family
\begin{align}
   p^{\boldsymbol{\hat{\theta}}}_{\boldsymbol{S}}(\boldsymbol{s})&= \exp (\hat{\theta}^1 s_1 + {\hat{\theta}}^{2} s_2 + {\hat{\theta}}^{12} s_1 s_2 - \phi_{\boldsymbol{S}} ({\hat{\theta}}^1, {\hat{\theta}}^2, {\hat{\theta}}^{12})  ), \nonumber\\
    \phi_{\boldsymbol{S}}({\hat{\theta}}^1, {\hat{\theta}}^2, {\hat{\theta}}^{12} )&= \ln \left[1+ \exp ({\hat{\theta}}^1)+\exp ({\hat{\theta}}^2)+\exp ({\hat{\theta}}^1+{\hat{\theta}}^2 + {\hat{\theta}}^{12}) \right],
\end{align}
which implies  
\begin{align}
p^{\boldsymbol{\hat{\theta}}}_{\boldsymbol{S}}(\boldsymbol{s})=
  \begin{cases}
   \exp (- \phi_{\boldsymbol{S}}({\hat{\theta}}^1, {\hat{\theta}}^2, {\hat{\theta}}^{12})  ) & (s_1=0, s_2=0), \\
   \exp ({\hat{\theta}}^{2}  - \phi_{\boldsymbol{S}}({\hat{\theta}}^1, {\hat{\theta}}^2, {\hat{\theta}}^{12})  ) & (s_1=0, s_2=1),\\
   \exp ({\hat{\theta}}^{1}  - \phi_{\boldsymbol{S}}({\hat{\theta}}^1, {\hat{\theta}}^2, {\hat{\theta}}^{12})  ) & (s_1=1, s_2=0), \\
   \exp ({\hat{\theta}}^{1} + {\hat{\theta}}^2 + {\hat{\theta}}^{12}  - \phi_{\boldsymbol{S}} ({\hat{\theta}}^1, {\hat{\theta}}^2, {\hat{\theta}}^{12})  )  & (s_1=1, s_2=1).
  \end{cases}
  \label{supspin2}
\end{align}
The transition probability $T(s_2,s_1) = p^{\boldsymbol{\hat{\theta}}}_{\boldsymbol{S}}(\boldsymbol{s})/ [\sum_{s_2} p^{\boldsymbol{\hat{\theta}}}_{\boldsymbol{S}}(\boldsymbol{s})]$ is given by
\begin{align}
   T(s_2,s_1) &= \exp ({\hat{\theta}}^2 s_2 + {\hat{\theta}}^{12} s_1 s_2 - \phi_{S_2|S_1} (s_1| {\hat{\theta}}^2, {\hat{\theta}}^{12})), \nonumber\\
    \phi_{S_2|S_1}(s_1| {\hat{\theta}}^2, {\hat{\theta}}^{12} )&= \ln \left[1+\exp ({\hat{\theta}}^2 + {\hat{\theta}}^{12}s_1) \right].
\end{align}
Because of one-to-one correspondence, we identify $p_{\boldsymbol{S}}(\boldsymbol{s})$ with $p^{\boldsymbol{\hat{\theta}}}_{\boldsymbol{S}}(\boldsymbol{s})$.
From Eqs.~(\ref{supspin}) and (\ref{supspin2}), we obtain the relationship between $({\hat{\theta}}^1, {\hat{\theta}}^2, {\hat{\theta}}^{12})$ and $(W(0 \to 1;t), W(1 \to 0;t), p (0;t))$ as
\begin{align}
   \phi_{\boldsymbol{S}}({\hat{\theta}}^1, {\hat{\theta}}^2, {\hat{\theta}}^{12}) 
   &= \ln \frac{1}{p_{\boldsymbol{S}}(0,0)} \nonumber  \\
   &= -\ln [(1-  W(0 \to 1;t)  dt) p (0;t)], \\
   {\hat{\theta}}^{1} &= \phi_{\boldsymbol{S}} ({\hat{\theta}}^1, {\hat{\theta}}^2, {\hat{\theta}}^{12})  + \ln[W(1 \to 0;t) dt (1-p (0;t))]\nonumber\\
   &=  \ln \frac{p_{\boldsymbol{S}} (1,0)}{p_{\boldsymbol{S}}(0,0)}\nonumber\\
   &=\ln \frac{W(1 \to 0;t) dt (1-p (0;t))}{(1-  W(0 \to 1;t)  dt) p (0;t)}, \\
   {\hat{\theta}}^{2} &= \phi_{\boldsymbol{S}} ({\hat{\theta}}^1, {\hat{\theta}}^2, {\hat{\theta}}^{12})  + \ln[W(0 \to 1;t) dtp (0;t)] \nonumber\\
    &=  \ln \frac{p_{\boldsymbol{S}}(0,1)}{p_{\boldsymbol{S}} (0,0)}  \nonumber\\
   &= \ln \frac{W(0 \to 1;t) dt}{1-  W(0 \to 1;t)  dt},          \\
   {\hat{\theta}}^{12} &= \phi_{\boldsymbol{S}}({\hat{\theta}}^1, {\hat{\theta}}^2, {\hat{\theta}}^{12})- {\hat{\theta}}^{1} -{\hat{\theta}}^{2}  + \ln[(1-  W(1 \to 0;t)  dt) (1-p (0;t))] \nonumber\\
   &=\ln \frac{p_{\boldsymbol{S}} (0,0)p_{\boldsymbol{S}} (1,1)}{p_{\boldsymbol{S}} (0,1)p_{\boldsymbol{S}} (1,0)}\nonumber\\
   &=  \ln \frac{[1-  W(0 \to 1;t)  dt][1-  W(1 \to 0;t)  dt] }{[W(0 \to 1;t) dt][W(1 \to 0;t) dt]}.
\end{align}

We here consider the backward manifold defined as
\begin{align}
\mathcal{M}_{\rm B} = \{ q_{\boldsymbol{S}}  |  q_{\boldsymbol{S}}({\boldsymbol{s}}) =  q_{S_2}(s_2)T(s_1, s_2) \}.
\end{align} 
If we use the expression of the exponential family for $ q_{\boldsymbol{S}}({\boldsymbol{s}}) =p^{\boldsymbol{{\theta}}}_{\boldsymbol{S}}(\boldsymbol{s})$, the reversible manifold is given by
\begin{align}
\mathcal{M}_{\rm B} = \{ p^{\boldsymbol{{\theta}}}_{\boldsymbol{S}}(\boldsymbol{s}) |  {\theta}^1=  {\hat{\theta}}^2,  {\theta}^{12}=  {\hat{\theta}}^{12}\},
\end{align} 
because the condition $q_{\boldsymbol{S}}({\boldsymbol{s}}) =  q_{S_2}(s_2)T(s_1|s_2)$ can be written as
\begin{align}
\exp ({{\theta}}^1 s_1 + {{\theta}}^{12} s_1 s_2 - \phi_{S_1|S_2} (s_2| {{\theta}}^1, {{\theta}}^{12})) &= \exp ({\hat{\theta}}^2 s_1 + {\hat{\theta}}^{12} s_2 s_1 - \phi_{S_2|S_1} (s_2| {\hat{\theta}}^2, {\hat{\theta}}^{12})),\\
\phi_{S_1|S_2}(s_1| {{\theta}}^1, {{\theta}}^{12} )&= \ln \left[1+\exp ({{\theta}}^1 + {{\theta}}^{12}s_1) \right].
\end{align} 

We here obtain the following Pythagorean theorem for any $q_{\boldsymbol{S}}  \in \mathcal{M}_{\rm B} $, 
\begin{align}
D(p_{\boldsymbol{S}} ||q_{\boldsymbol{S}}  )&=D(p_{\boldsymbol{S}}  ||q^*_{\boldsymbol{S}} )+ D(q^*_{\boldsymbol{S}} ||q_{\boldsymbol{S}} ), \nonumber\\
q^*_{\boldsymbol{S}} ({\boldsymbol{s}} ) &=   \exp (\hat{\theta}^2 s_1 + {\theta}^{2*} s_2 + \hat{\theta}^{12} s_1 s_2 - \phi_{\boldsymbol{S}} (\hat{\theta}^2, {\theta}^{2*}, \hat{\theta}^{12})  ), 
\end{align} 
with the constraint
\begin{align}
\sum_{s_1} q^*_{\boldsymbol{S}}({\boldsymbol{s}}) &= \sum_{s_1} p_{\boldsymbol{S}}({\boldsymbol{s}}).
\label{supconst}
\end{align} 
In our main result, the total entropy production is given by the following optimization problem 
\begin{align}
\sigma_{\rm tot}^{\mathcal{Z}}=D^{\rm opt}(p_{\boldsymbol{S}}||\mathcal{M}_{\rm B}) = D(p_{\boldsymbol{S}}  ||q^*_{\boldsymbol{S}} ).
\end{align} 
By using the expression by $({\theta}_1,{\theta}_2,{\theta}_{12})$, this optimization problem can be written as 
\begin{align}
\sigma_{\rm tot}^{\mathcal{Z}} &= {\rm min}_{q_{\boldsymbol{S}}  \in \mathcal{M}_{\rm B}} D(p_{\boldsymbol{S}}  ||q_{\boldsymbol{S}})\\
&={\rm min}_{{\theta}^2} [\mathbb{E}[s_1]({\hat{\theta}}^1-{\hat{\theta}}^2) +\mathbb{E}[s_2] ({\hat{\theta}}^2 - {\theta}^2) -  \phi_{\boldsymbol{S}}({\hat{\theta}}^1, {\hat{\theta}}^2, {\hat{\theta}}^{12}) + \phi_{\boldsymbol{S}}({\hat{\theta}}^2, {\theta}^2, {\hat{\theta}}^{12}) ] \\
&=\mathbb{E}[s_1]({\hat{\theta}}^1-{\hat{\theta}}^2) +\mathbb{E}[s_2] ({\hat{\theta}}^2 - {\theta}^{2*}) -  \phi_{\boldsymbol{S}}({\hat{\theta}}^1, {\hat{\theta}}^{2*}, {\hat{\theta}}^{12}) + \phi_{\boldsymbol{S}}({\hat{\theta}}^2, {\theta}^{2*}, {\hat{\theta}}^{12}),
\label{supopt}
\end{align} 
where $\mathbb{E}$ denotes the expected value $\mathbb{E}[\cdots] = \sum_{{\boldsymbol{s}}}p_{\boldsymbol{S}}({\boldsymbol{s}}) \cdots$.
The constraint Eq. (\ref{supconst}) is calculated as
\begin{align}
\exp \left[({\hat{\theta}}^2 - {\theta}^{2*} )s_2 - \phi_{\boldsymbol{S}}({\hat{\theta}}^1, {\hat{\theta}}^2, {\hat{\theta}}^{12}) +\phi_{\boldsymbol{S}}({\hat{\theta}}^2, {\theta}^{2*}, {{\hat{\theta}}}^{12}) \right] &= \exp \left[ \phi_{S_1|S_2}(s_2|{\hat{\theta}}^2,  {\hat{\theta}}^{12} )-\phi_{S_1|S_2}(s_2|{\hat{\theta}}^1, {\hat{\theta}}^{12} ) \right].
\label{supconstAA}
\end{align} 
Under the constraint Eq.~(\ref{supconstAA}), the optimization problem Eq. (\ref{supopt}) is calculated as
\begin{align}
\sigma_{\rm tot}^{\mathcal{Z}}&= \mathbb{E} \left[s_1({\hat{\theta}}^1-{\hat{\theta}}^2) +s_2({\hat{\theta}}^2 - {\theta}^{2*}) -  \phi_{\boldsymbol{S}} ({\hat{\theta}}^1, {\hat{\theta}}^2, {\hat{\theta}}^{12}) + \phi_{\boldsymbol{S}} ({\hat{\theta}}^2, {\theta}^{2*}, {\hat{\theta}}^{12}) \right] \nonumber\\
&= \mathbb{E} \left[s_1({\hat{\theta}}^1-{\hat{\theta}}^2) + \phi_{S_1|S_2}(s_2|{\hat{\theta}}^2,  {\hat{\theta}}^{12} )-\phi_{S_1|S_2}(s_2|{\hat{\theta}}^1, {\hat{\theta}}^{12} )\right].
\label{supen}
\end{align} 
We can check the equivalence between Eq. (\ref{supen}) and the original definition of the total entropy production as follows,
\begin{align}
\sigma_{\rm tot}^{\mathcal{Z}}&= \sum_{\boldsymbol{s}} T(s_2,s_1)p_{S_1}(s_1) \ln \frac{T(s_2,s_1)p_{S_1}(s_1)  }{T(s_1,s_2)p_{S_2}(s_2) }  \nonumber \\
&=  \mathbb{E} \left[\ln \frac{T(s_2,s_1)p_{S_1}(s_1)}{T(s_1|s_2)p_{S_2}(s_2)} \right] \nonumber \\
&=  \mathbb{E} \left[\ln \frac{p_{S_1|S_2}(s_1|s_2)}{\exp ({\hat{\theta}}^2 s_1 + {\hat{\theta}}^{12} s_1 s_2 - \phi_{S_2|S_1} (s_2| {\hat{\theta}}^2, {\hat{\theta}}^{12})) } \right] \nonumber \\
&=  \mathbb{E} \left[\ln \frac{\exp ({\hat{\theta}}^1 s_1 + {\hat{\theta}}^{12} s_1 s_2 - \phi_{S_1|S_2} (s_2| {\hat{\theta}}^2, {\hat{\theta}}^{12}))}{\exp ({\hat{\theta}}^2 s_1 + {\hat{\theta}}^{12} s_1 s_2- \phi_{S_2|S_1} (s_2| {\hat{\theta}}^2, {\hat{\theta}}^{12})) } \right]\nonumber  \\
&= \mathbb{E} \left[s_1 ({\hat{\theta}}^1-{\hat{\theta}}^2) + \phi_{S_1|S_2}(s_2|{\hat{\theta}}^2,  {\hat{\theta}}^{12} )-\phi_{S_1 |S_2}(s_2|{\hat{\theta}}^1, {\hat{\theta}}^{12} )\right],
\end{align} 
where we used $ \phi_{S_2|S_1} (s_2| {\hat{\theta}}^2, {\hat{\theta}}^{12})= \phi_{S_1|S_2} (s_2| {\hat{\theta}}^2, {\hat{\theta}}^{12})$.

\subsection{III. The detailed calculation of Example II: Two spins model}
We start with the joint distribution
\begin{align}  
 p_{\boldsymbol{S}}^{\hat{\boldsymbol{\theta}}}(\boldsymbol{s})=& \exp \left[ \sum_{i} s_i \hat{\theta}^i +\sum_{i<j} s_i s_j \hat{\theta}^{ij} + \sum_{i<j<k} s_i s_j s_k \hat{\theta}^{ijk}  +\sum_{i<j<k<l} s_i s_j s_ks_l \hat{\theta}^{ijkl}  - \phi_{\boldsymbol{S}} (\boldsymbol{\hat{\theta}}) \right],
\end{align} 
where $\boldsymbol{s}=(s_1,s_2, s_3,s_4)=(x,y,x',y')$ is the spin notation with $s_i \in \{0,1\}$, and $\phi_{\boldsymbol{S}}  (\boldsymbol{\hat{\theta}})$ is the normalization constant that satisfies $\sum_{\boldsymbol{s}}p_{\boldsymbol{S}}^{\hat{\boldsymbol{\theta}}}(\boldsymbol{s})= 1$. 

We consider the both bipartite conditions $\mathcal{C}_{\rm BI}$ and $\mathcal{C}_{\rm BI}^*$. We here compare $p^{\hat{\boldsymbol{\theta}}}_{{\boldsymbol X'}|{\boldsymbol Z}}({\boldsymbol x'}|{\boldsymbol z})= \sum_{s_4} p_{\boldsymbol{S}}^{\hat{\boldsymbol{\theta}}}(\boldsymbol{s}) / [ \sum_{s_3, s_4} p_{\boldsymbol{S}}^{\hat{\boldsymbol{\theta}}}(\boldsymbol{s}) ]$ with $p^{\hat{\boldsymbol{\theta}}}_{{\boldsymbol X'}|{\boldsymbol Z},{\boldsymbol Y'} }({\boldsymbol x'}|{\boldsymbol z},{\boldsymbol y'}) =p_{\boldsymbol{S}}^{\hat{\boldsymbol{\theta}}}(\boldsymbol{s}) / [ \sum_{s_3} p_{\boldsymbol{S}}^{\hat{\boldsymbol{\theta}}}(\boldsymbol{s}) ]$. The conditional probability $p^{\hat{\boldsymbol{\theta}}}_{{\boldsymbol X'}|{\boldsymbol Z}}({\boldsymbol x'}|{\boldsymbol z})$ is calculated as
\begin{align}  
\ln p^{\hat{\boldsymbol{\theta}}}_{{\boldsymbol X'}|{\boldsymbol Z}}({\boldsymbol x'}|{\boldsymbol z})=& s_3 \hat{\theta}^3 +s_1 s_3 \hat{\theta}^{13} +s_2 s_3 \hat{\theta}^{23} + s_1 s_2 s_3 \hat{\theta}^{123}- \phi_{{\boldsymbol X'}|{\boldsymbol Z}} (s_1,s_2|\boldsymbol{\hat{\theta}}), \nonumber \\
\phi_{{\boldsymbol X'}|{\boldsymbol Z}} (s_1,s_2|\boldsymbol{\hat{\theta}}) :=& \ln \left[\exp \left(\hat{\theta}^3 + s_1 \hat{\theta}^{13}+ s_2 \hat{\theta}^{23} + s_1 s_2 \hat{\theta}^{123}  \right) +1 \right].
\label{bi1}
\end{align} 
The conditional probability $p^{\hat{\boldsymbol{\theta}}}_{{\boldsymbol X'}|{\boldsymbol Z},{\boldsymbol Y'} }({\boldsymbol x'}|{\boldsymbol z},{\boldsymbol y'})$ is calculated as 
\begin{align}  
\ln p^{\hat{\boldsymbol{\theta}}}_{{\boldsymbol X'}|{\boldsymbol Z},{\boldsymbol Y'} }({\boldsymbol x'}|{\boldsymbol z},{\boldsymbol y'}) =& s_3 \hat{\theta}^3 +s_1 s_3 \hat{\theta}^{13} +s_2 s_3 \hat{\theta}^{23} +s_3 s_4 \hat{\theta}^{34} + s_1 s_2 s_3 \hat{\theta}^{123} \nonumber \\
&+s_1 s_3 s_4 \hat{\theta}^{134} +
s_2 s_3 s_4 \hat{\theta}^{234} +s_1 s_2 s_3 s_4 \hat{\theta}^{1234}- \phi_{{\boldsymbol X'}|{\boldsymbol Z},{\boldsymbol Y'}} (s_1,s_2,s_4|\boldsymbol{\hat{\theta}}), \nonumber \\
\phi_{{\boldsymbol X'}|{\boldsymbol Z},{\boldsymbol Y'}} (s_1,s_2,s_4|\boldsymbol{\hat{\theta}}) :=& \ln \left[\exp \left(\hat{\theta}^3 + s_1 \hat{\theta}^{13} + s_2 \hat{\theta}^{23}+s_4 \hat{\theta}^{34}+ s_1 s_2 \hat{\theta}^{123}+s_1  s_4 \hat{\theta}^{134} +s_2 s_4 \hat{\theta}^{234} +s_1 s_2 s_4 \hat{\theta}^{1234} \right) +1 \right].
\label{bi2}
\end{align} 
From Eqs. (\ref{bi1}) and (\ref{bi2}), we obtain the condition of $\mathcal{C}_{\rm BI} :p^{\hat{\boldsymbol{\theta}}}_{{\boldsymbol X'}|{\boldsymbol Z},{\boldsymbol Y'} } =p^{\hat{\boldsymbol{\theta}}}_{{\boldsymbol X'}|{\boldsymbol Z}}$ as
\begin{align}  
\mathcal{C}_{\rm BI}: \hat{\theta}^{34}=\hat{\theta}^{134}=\hat{\theta}^{234}= \hat{\theta}^{1234}=0.
\end{align} 
In the same way, we also obtain the condition of $\mathcal{C}_{\rm BI}^*$ as
\begin{align}  
\mathcal{C}_{\rm BI}^*: \hat{\theta}^{12}=\hat{\theta}^{123}=\hat{\theta}^{124}= \hat{\theta}^{1234}=0.
\end{align} 
To clarify the relationship between $\mathcal{C}_{\rm BI}$ and $\mathcal{C}_{\rm BI}^*$, we can consider the permutation $(\alpha(1),\alpha(2),\alpha(3),\alpha(4)) = (3,4,1,2)$. The condition of $\mathcal{C}_{\rm BI}^*$ is given by the condition of $\mathcal{C}_{\rm BI}$ with the permutation $\alpha$,
\begin{align}  
\mathcal{C}_{\rm BI}^*: \hat{\theta}^{\alpha(3)\alpha(4)}=\hat{\theta}^{\alpha(3)\alpha(4)\alpha(1)}=\hat{\theta}^{\alpha(3)\alpha(4)\alpha(2)}= \hat{\theta}^{\alpha(3)\alpha(4)\alpha(1)\alpha(2)}=0.
\end{align} 

Next, we discuss the backward manifold $\mathcal{M}_{\rm B}$. The transition probability $T({\boldsymbol z'}, {\boldsymbol z})=p^{\hat{\boldsymbol{\theta}}}_{{\boldsymbol Z'}|{\boldsymbol Z}}({\boldsymbol z'}|{\boldsymbol z})= p_{\boldsymbol{S}}^{\hat{\boldsymbol{\theta}}}(\boldsymbol{s}) / [ \sum_{s_3, s_4} p_{\boldsymbol{S}}^{\hat{\boldsymbol{\theta}}}(\boldsymbol{s}) ]$ is calculated as
\begin{align}  
\ln T ({\boldsymbol z'}, {\boldsymbol z})=& s_3 \hat{\theta}^3 + s_4 \hat{\theta}^4 +\sum_{i<4} s_i s_4 \hat{\theta}^{i4}+\sum_{i<3} s_i s_3 \hat{\theta}^{i3}  \nonumber \\
&+ \sum_{i<j<k} s_i s_j s_k \hat{\theta}^{ijk}+\sum_{i<j<k<l} s_i s_j s_k s_l \hat{\theta}^{ijkl}  - \phi_{{\boldsymbol Z'}|{\boldsymbol Z}} (s_1, s_2| \boldsymbol{\hat{\theta}} ), \nonumber
\end{align} 
\begin{align}  
 &\phi_{{\boldsymbol Z'}|{\boldsymbol Z}} (s_1, s_2|\boldsymbol{\hat{\theta}} ) &\nonumber \\
& := \ln \left[\exp (\hat{\theta}^3 + \hat{\theta}^4 +s_1 \hat{\theta}^{14}+s_2 \hat{\theta}^{24}+\hat{\theta}^{34} +s_1 \hat{\theta}^{13} +s_2 \hat{\theta}^{23} +  s_1 s_2 \hat{\theta}^{123}+  s_1 s_2 \hat{\theta}^{124} +  s_1 \hat{\theta}^{134} +   s_2 \hat{\theta}^{234}  +s_1 s_2 \hat{\theta}^{1234}  ) \right. \nonumber \\
&\left. +\exp (\hat{\theta}^3 +s_1 \hat{\theta}^{13} +s_2 \hat{\theta}^{23} +  s_1 s_2 \hat{\theta}^{123}) +\exp (\hat{\theta}^4 +s_1 \hat{\theta}^{14}+s_2 \hat{\theta}^{24} +  s_1 s_2 \hat{\theta}^{124} )+1 \right].
\label{sup3}
\end{align} 
The conditional probability $p^{{\boldsymbol{\hat{\theta}}}}_{{\boldsymbol Z}|{\boldsymbol Z'}}({\boldsymbol z}|{\boldsymbol z'})= p_{\boldsymbol{S}}^{\hat{\boldsymbol{\theta}}}(\boldsymbol{s}) / [ \sum_{s_1, s_2} p_{\boldsymbol{S}}^{\hat{\boldsymbol{\theta}}}(\boldsymbol{s}) ]$ is also calculated as
\begin{align}  
\ln p^{{\boldsymbol{\hat{\theta}}}}_{{\boldsymbol Z}|{\boldsymbol Z'}}({\boldsymbol z}|{\boldsymbol z'}) =& s_1 \hat{\theta}^1 + s_2 \hat{\theta}^2 +\sum_{1<i} s_1 s_i \hat{\theta}^{1i}+\sum_{2<i} s_2 s_i \hat{\theta}^{2i}  \nonumber \\
&+ \sum_{i<j<k} s_i s_j s_k \hat{\theta}^{ijk}+\sum_{i<j<k<l} s_i s_j s_ks_l \hat{\theta}^{ijkl}  - \phi_{{\boldsymbol Z}|{\boldsymbol Z'}} (s_3, s_4| {\boldsymbol{ \hat{\theta}}}), \nonumber
\end{align} 
\begin{align}  
 & \phi_{{\boldsymbol Z}|{\boldsymbol Z'}} (s_3, s_4| {\boldsymbol{ \hat{\theta}}}) &\nonumber \\
& := \ln \left[\exp (\hat{\theta}^1 + \hat{\theta}^2 +s_3 \hat{\theta}^{23}+s_4 \hat{\theta}^{24}+\hat{\theta}^{12} +s_3 \hat{\theta}^{13} +s_4 \hat{\theta}^{14} +  s_3 s_4\hat{\theta}^{134}+  s_3 s_4 \hat{\theta}^{234} +  s_3 \hat{\theta}^{123} +   s_4 \hat{\theta}^{124}  +s_3 s_4 \hat{\theta}^{1234}  ) \right. \nonumber \\
&\left. +\exp (\hat{\theta}^1 +s_3 \hat{\theta}^{13} +s_4 \hat{\theta}^{14} +  s_3s_4 \hat{\theta}^{134}) +\exp (\hat{\theta}^2 +s_3 \hat{\theta}^{23}+s_4 \hat{\theta}^{24} +  s_3 s_4 \hat{\theta}^{234} +  s_3 \hat{\theta}^{123} )+1 \right].
\label{sup4}
\end{align} 
The backward manifold is defined as 
\begin{align}
\mathcal{M}_{\rm B}= \{   p_{\boldsymbol{S}}^{{\boldsymbol{\theta}}}(\boldsymbol{s})   |  p_{\boldsymbol{S}}^{{\boldsymbol{\theta}}}(\boldsymbol{s}) = p^{{\boldsymbol{{\theta}}}}_{{\boldsymbol Z'}}({\boldsymbol z'})T({\boldsymbol z}, {\boldsymbol z'})  \},
\end{align}
where $ p^{{\boldsymbol{{\theta}}}}_{{\boldsymbol Z'}}({\boldsymbol z'})= \sum_{{\boldsymbol z}}  p_{\boldsymbol{S}}^{{\boldsymbol{\theta}}}(\boldsymbol{s})$. The equations (\ref{sup3}) and (\ref{sup4}) yield 
\begin{align}
\mathcal{M}_{\rm B}= \left\{ p_{\boldsymbol{S}}^{\hat{\boldsymbol{\theta}}} \left| \theta^1=\hat{\theta}^3, \:\theta^2=\hat{\theta}^4, \:\theta^{23}=\hat{\theta}^{14},  \: \theta^{24}= \hat{\theta}^{24},  \: \theta^{12}= \hat{\theta}^{34},    
\: \theta^{13}= \hat{\theta}^{13},   \: \theta^{14}= \hat{\theta}^{23},\right.  \right. \nonumber \\ \: \left. \theta^{134}= \hat{\theta}^{123}, \: \theta^{234}= \hat{\theta}^{124}, \: \theta^{123}= \hat{\theta}^{134}, \: \theta^{124}= \hat{\theta}^{234}, \: \theta^{1234}= \hat{\theta}^{1234} \right\}.
\end{align} 
Under the both bipartite conditions $\mathcal{C}_{\rm BI}$ and $\mathcal{C}_{\rm BI}^*$, the joint probability is given by 
\begin{align}  
{p^{\rm BI}}_{\boldsymbol{S}}^{ \hat{\boldsymbol{\theta}}}= \left.  p_{\boldsymbol{S}}^{\hat{\boldsymbol{\theta}}} \right|_{\hat{\theta}^{34}=\hat{\theta}^{134}=\hat{\theta}^{234}=  \hat{\theta}^{12}=\hat{\theta}^{123}=\hat{\theta}^{124}= \hat{\theta}^{1234}=0}.
\end{align}
For this distribution ${p^{\rm BI}}_{\boldsymbol{S}}^{ \hat{\boldsymbol{\theta}}}$, the condition of the backward manifold is given by
\begin{align}
\mathcal{M}_{\rm B}= \left\{ {p^{\rm BI}}_{\boldsymbol{S}}^{ {\boldsymbol{\theta}}} \left| \theta^1=\hat{\theta}^3, \:\theta^2=\hat{\theta}^4, \:\theta^{23}=\hat{\theta}^{14},  \: \theta^{24}= \hat{\theta}^{24},    
\: \theta^{13}= \hat{\theta}^{13},   \: \theta^{14}= \hat{\theta}^{23}\right.  \right\}. 
\end{align} '

Next, we discuss the local backward manifold $\mathcal{M}^{\mathcal{X}}_{\rm LB}$.  Then the transition probability $T^{\mathcal{X}}( {\boldsymbol z'},{\boldsymbol z})= p^{{\boldsymbol{{\theta}}}}_{{\boldsymbol X'}|{\boldsymbol Z}, {\boldsymbol Y'}}({\boldsymbol x'}|{\boldsymbol z})$ is given by Eq. (\ref{bi2}). The conditional probability $p^{\hat{\boldsymbol{\theta}}}_{{\boldsymbol X}|{\boldsymbol Z'},{\boldsymbol Y} }({\boldsymbol x}|{\boldsymbol z'},{\boldsymbol y}) =  p_{\boldsymbol{S}}^{\hat{\boldsymbol{\theta}}}(\boldsymbol{s}) / [ \sum_{s_1} p_{\boldsymbol{S}}^{\hat{\boldsymbol{\theta}}}(\boldsymbol{s}) ]$ is calculated as
\begin{align}  
\ln p^{\hat{\boldsymbol{\theta}}}_{{\boldsymbol X}|{\boldsymbol Z'},{\boldsymbol Y} }({\boldsymbol x}|{\boldsymbol z'},{\boldsymbol y}) =& s_1 \hat{\theta}^1 +s_1 s_2 \hat{\theta}^{12} +s_1 s_3 \hat{\theta}^{13} +s_1 s_4 \hat{\theta}^{14} + s_1 s_2 s_3 \hat{\theta}^{123} \nonumber \\
&+s_1 s_2 s_4 \hat{\theta}^{124} +
s_1 s_3 s_4 \hat{\theta}^{134} +s_1 s_2 s_3 s_4 \hat{\theta}^{1234}- \phi_{{\boldsymbol X}|{\boldsymbol Z'},{\boldsymbol Y}} (s_2,s_3,s_4|\boldsymbol{\hat{\theta}}), \nonumber \\
\phi_{{\boldsymbol X}|{\boldsymbol Z'},{\boldsymbol Y}} (s_2,s_3,s_4|\boldsymbol{\hat{\theta}}) :=& \ln \left[\exp \left(\hat{\theta}^1 + s_2 \hat{\theta}^{12} + s_3 \hat{\theta}^{13}+s_4 \hat{\theta}^{14}+ s_2 s_3 \hat{\theta}^{123}+s_2  s_4 \hat{\theta}^{124} +s_3 s_4 \hat{\theta}^{134} +s_2 s_3 s_4 \hat{\theta}^{1234} \right) +1 \right].
\label{sup6}
\end{align} 
The local backward manifold is defined as 
\begin{align}
\mathcal{M}^{\mathcal{X}}_{\rm LB}= \{p^{{\boldsymbol{{\theta}}}}_{\boldsymbol{S}} | p^{{\boldsymbol{{\theta}}}}_{\boldsymbol{S}} (\boldsymbol{s})=   p^{{\boldsymbol{{\theta}}}}_{\boldsymbol{Z'},\boldsymbol{Y} }(\boldsymbol{z'},\boldsymbol{y} ) T^{\mathcal{X}}({\boldsymbol z}, {\boldsymbol z'}) \},
\end{align}
where $p^{{\boldsymbol{{\theta}}}}_{\boldsymbol{Z'},\boldsymbol{Y} }(\boldsymbol{z'},\boldsymbol{y} ) = \sum_{s_1} p^{{\boldsymbol{{\theta}}}}_{\boldsymbol{S}} (\boldsymbol{s})$. 
The equations (\ref{bi2}) and (\ref{sup6}) yield 
\begin{align}
\mathcal{M}_{\rm LB}^{\mathcal X}= \left\{p^{{\boldsymbol{{\theta}}}}_{\boldsymbol{S}} \left| \theta^1=\hat{\theta}^3,  \:\theta^{12}=\hat{\theta}^{34}, \:\theta^{13}=\hat{\theta}^{13}, \:\theta^{14}=\hat{\theta}^{23} , \: \theta^{123}= \hat{\theta}^{134}, \: \theta^{124}= \hat{\theta}^{234},  \right.  \right. \nonumber \\
\left. \left.  \theta^{134}= \hat{\theta}^{123}, \: \theta^{234}= \hat{\theta}^{124}, \: \theta^{1234}= \hat{\theta}^{1234} \right\} \right. .
\end{align} 
In the same way, we obtain the condition of $\mathcal{M}_{\rm LB}^{\mathcal Y}$
\begin{align}
\mathcal{M}_{\rm LB}^{\mathcal Y}= \left\{p^{{\boldsymbol{{\theta}}}}_{\boldsymbol{S}} \left| \theta^2=\hat{\theta}^4,  \:\theta^{12}=\hat{\theta}^{34}, \:\theta^{23}=\hat{\theta}^{14} , \:\theta^{24}=\hat{\theta}^{24},\: \theta^{123}= \hat{\theta}^{134},  \: \theta^{124}= \hat{\theta}^{234},  \right.  \right.\nonumber \\
\left. \left.  \: \theta^{134}= \hat{\theta}^{123}, \: \theta^{234}= \hat{\theta}^{124} , \: \theta^{1234}= \hat{\theta}^{1234} \right\} \right. .
\end{align} 
To clarify the relationship between $\mathcal{M}_{\rm LR}^{\mathcal X}$ and $\mathcal{M}_{\rm LR}^{\mathcal Y}$, we can consider the permutation $(\alpha'(1),\alpha'(2),\alpha'(3),\alpha'(4)) = (2,1,4,3)$. The condition of $\mathcal{M}_{\rm LR}^{\mathcal Y}$ is given by the condition of $\mathcal{M}_{\rm LR}^{\mathcal X}$ with the permutation $\alpha'$,
\begin{align}
\mathcal{M}_{\rm LB}^{\mathcal Y}= \left\{p^{{\boldsymbol{{\theta}}}}_{\boldsymbol{S}} \left| \theta^{\alpha'(1)
}=\hat{\theta}^{\alpha'(3)},  \:\theta^{\alpha'(1)\alpha'(2)}=\hat{\theta}^{\alpha'(3)\alpha'(4)}, \:\theta^{\alpha'(1)\alpha'(3)}=\hat{\theta}^{\alpha'(1)\alpha'(3)}, \:\theta^{\alpha'(1)\alpha'(4)}=\hat{\theta}^{\alpha'(2)\alpha'(3)} ,\right.  \right.\nonumber \\
\left. \left.  \: \theta^{\alpha'(1)\alpha'(2)\alpha'(3)}= \hat{\theta}^{\alpha'(1)\alpha'(3)\alpha'(4)}, \: \theta^{\alpha'(1)\alpha'(2)\alpha'(4)}= \hat{\theta}^{\alpha'(2)\alpha'(3)\alpha'(4)},   \theta^{\alpha'(1)\alpha'(3)\alpha'(4)}= \hat{\theta}^{\alpha'(1)\alpha'(2)\alpha'(3)}, \right.  \right.\nonumber \\
\left. \left. 
 \: \theta^{\alpha'(2)\alpha'(3)\alpha'(4)}= \hat{\theta}^{\alpha'(1)\alpha'(2)\alpha'(4)}, \: \theta^{\alpha'(1)\alpha'(2)\alpha'(3)\alpha'(4)}= \hat{\theta}^{\alpha'(1)\alpha'(2)\alpha'(3)\alpha'(4)} \right\} \right. .
\end{align} 
For this distribution ${p^{\rm BI}}_{\boldsymbol{S}}^{ \hat{\boldsymbol{\theta}}}$ under the both bipartite conditions, the local backward manifolds are given by
\begin{align}
\mathcal{M}_{\rm LB}^{\mathcal X}= \left\{ {p^{\rm BI}}_{\boldsymbol{S}}^{ {\boldsymbol{\theta}}} \left|  \theta^1=\hat{\theta}^3,  \:\theta^{13}=\hat{\theta}^{13}, \:\theta^{14}=\hat{\theta}^{23} \right\} \right.,\\
\mathcal{M}_{\rm LB}^{\mathcal Y}= \left\{ {p^{\rm BI}}_{\boldsymbol{S}}^{ {\boldsymbol{\theta}}} \left| \theta^2=\hat{\theta}^4,  \:\theta^{24}=\hat{\theta}^{24}, \:\theta^{23}=\hat{\theta}^{14}  \right\} \right..
\end{align} 

\subsection{IV. The case of feedback control} We consider the situation that the time evolution of the system $\mathcal{X}$ depends on the fixed memory $\mathcal{M}$. This situation is well known as the problem of the Maxwell's demon under feedback control. We show that the partial entropy production for this case can also be discussed in our unified framework.

Let $\boldsymbol{X}$ and $\boldsymbol{X'}$ be random variables of the system $\mathcal{X}$ at time $t$ and $t+dt$, respectively. 
Let $\boldsymbol{M}$ be a random variable of the memory $\mathcal{M}$. We denotes the set of random variables as $\boldsymbol{S} = \{\boldsymbol{X}, \boldsymbol{X'},\boldsymbol{M}\}$, and the set of states as $\boldsymbol{s} = \{\boldsymbol{x}, \boldsymbol{x'},\boldsymbol{m}\}$, respectively. The joint probability of $\boldsymbol{S}$ is given by $p_{\boldsymbol{S}}(\boldsymbol{s})$. We consider the situation that the transition probability of $\mathcal{X}$ depend on the state of memory,
\begin{align}
p_{\boldsymbol{X'}|\boldsymbol{X}, \boldsymbol{M}}(\boldsymbol{x'}|\boldsymbol{x}, \boldsymbol{m})  =: T^{\mathcal{XM}}(\boldsymbol{x'}, \boldsymbol{m},\boldsymbol{x}),
\end{align} 
where $p_{\boldsymbol{X'}|\boldsymbol{X}, \boldsymbol{M}}(\boldsymbol{x'}|\boldsymbol{x}, \boldsymbol{m})  = p_{\boldsymbol{S}}(\boldsymbol{s}) / [ \sum_{\boldsymbol{x'}} p_{\boldsymbol{S}}(\boldsymbol{s}) ]$. We here introduce {\it the feedback backward manifold} such that
\begin{align}
\mathcal{M}_{\rm FB}= \{q_{\boldsymbol{S}} | q_{\boldsymbol{S}} (\boldsymbol{s}) = T^{\mathcal{XM}} (\boldsymbol{x}, \boldsymbol{m},\boldsymbol{x'}) q_{\boldsymbol{X'} \boldsymbol{M}} (\boldsymbol{x'}, \boldsymbol{m}) \},
\end{align}
where $q_{\boldsymbol{X'} \boldsymbol{M}} (\boldsymbol{x'}, \boldsymbol{m})= \sum_{\boldsymbol{x}} q_{\boldsymbol{S}} (\boldsymbol{s})$.
The feedback reversible manifold is equivalent to the reversible manifold $\mathcal{M}_{\rm B} = \mathcal{M}_{\rm FB}$, if we consider the time evolution from  $\boldsymbol{Z} = \{\boldsymbol{X}, \boldsymbol{M} \}$ to $\boldsymbol{Z'} = \{\boldsymbol{X'}, \boldsymbol{M}\}$. If the joint probability $q_{\boldsymbol{S}}$ is on this manifold $\mathcal{M}_{\rm FB}$, dynamics of $\mathcal{X}$ are reversible in time under feedback control. If we introduce the joint probability $q^{\mathcal{XM}*}_{\boldsymbol{S}} (\boldsymbol{s})= T^{\mathcal{XM}} (\boldsymbol{x}, \boldsymbol{m},\boldsymbol{x'}) p_{ \boldsymbol{X'}, \boldsymbol{M}}( \boldsymbol{x'}, \boldsymbol{m})$, the following Pythagorean theorem is valid for any $q_{\boldsymbol{S}} \in \mathcal{M}_{\rm FR}$,
\begin{align}
D(p_{\boldsymbol{S}}||q_{\boldsymbol{S}}) = D(p_{\boldsymbol{S}}||q^{\mathcal{XM}*}_{\boldsymbol{S}} ) + D(q^{\mathcal{XM}*}_{\boldsymbol{S}}||q_{\boldsymbol{S}} ).
\label{Pythagorean}
\end{align} 
Thus, the feedback backward manifold is flat, and the solution of the optimization problem $D^{\rm opt}(p_{\boldsymbol{S}}||\mathcal{M}_{\rm FB})$ is given by
\begin{align}
D^{\rm opt}(p_{\boldsymbol{S}}||\mathcal{M}_{\rm FB}) &:= {\rm min}_{q_{\boldsymbol{S}} \in \mathcal{M}_{\rm FB}} D(p_{\boldsymbol{S}}||q_{\boldsymbol{S}})  \\
&= D(p_{\boldsymbol{S}}||q^{\mathcal{XM}*}_{\boldsymbol{S}} ).
\end{align} 

We here derive the result that the partial entropy production under feedback control $\sigma_{\rm feedback}^{\mathcal{X}}$ is given by the optimization problem
\begin{align}
\sigma_{\rm feedback}^{\mathcal{X}} = D^{\rm opt}(p_{\boldsymbol{S}}||\mathcal{M}_{\rm FB}).
\label{supinfofeed}
\end{align} 
The partial entropy production under feedback control $\sigma_{\rm feedback}^{\mathcal{X}}$ is defined as
\begin{align}
&\sigma_{\rm feedback}^{\mathcal{X}} := \sigma^{\mathcal{X}}_{\rm sys}+ \sigma^{\mathcal{X}}_{\rm bath}- \Delta \mathcal{I},\\
&\sigma^{\mathcal{X}}_{\rm sys} := H(\boldsymbol{X'})-H(\boldsymbol{X}),\\
&\sigma^{\mathcal{X}}_{\rm bath} := \mathbb{E} \left[ \ln \frac{T^{\mathcal{XM}}(\boldsymbol{x'},\boldsymbol{m}, \boldsymbol{x})}{T^{\mathcal{XM}}(\boldsymbol{x}, \boldsymbol{m}, \boldsymbol{x'})} \right], \\
&\Delta \mathcal{I} := I(\boldsymbol{X'};\boldsymbol{M} ) -I(\boldsymbol{X};\boldsymbol{M}),
\end{align} 
where $\sigma^{\mathcal{X}}_{\rm sys} $ is the entropy change of the system $\mathcal{X}$, $\sigma^{\mathcal{X}}_{\rm bath} $ is the entropy change of the heat bath attached to the system $\mathcal{X}$ and $\Delta  \mathcal{I}$ is the mutual information change between the system $\mathcal{X}$ and the memory $\mathcal{M}$. To show the following relationship
\begin{align}
\sigma_{\rm feedback}^{\mathcal{X}} = D(p_{\boldsymbol{S}}||q^{\mathcal{XM}*}_{\boldsymbol{S}} ),
\end{align} 
we obtain the result Eq.~(\ref{supinfofeed}). The second law of information thermodynamics under feedback control is given by the nonnegativity of $\sigma_{\rm feedback}^{\mathcal{X}}$,
\begin{align}
\sigma^{\mathcal{X}}_{\rm sys}+ \sigma^{\mathcal{X}}_{\rm bath} \geq \Delta \mathcal{I}.
\end{align}
This inequality implies the trade-off relationship between the entropy changes in the system $\mathcal{X}$ and the information between the system $\mathcal{X}$ and the memory $\mathcal{M}$.

\end{document}